\documentclass{aa}  
\usepackage{graphicx}
\usepackage{amsmath}
\usepackage{amssymb}
\usepackage{mathtools}
\usepackage{mathrsfs}
\usepackage{txfonts}
\newcommand\Mstar{M_*}
\newcommand{\rhos}{\rho_*}
\newcommand\Msun{M_{\odot}}
\newcommand\rh{r_{1/2}}
\newcommand\mk{m_k}
\newcommand{\xk}{x_k}
\newcommand{\yk}{y_k}
\newcommand{\zk}{z_k}

\newcommand{\vx}{v_x}
\newcommand{\vy}{v_y}
\newcommand{\vz}{v_z}
\newcommand{\vi}{v_i}
\newcommand{\vj}{v_j}
\newcommand{\vki}{v_{ki}}
\newcommand{\vkj}{v_{kj}}
\newcommand{\vR}{v_R}
\newcommand{\vphi}{v_{\varphi}}
\newcommand{\Deltac}{\Delta_R}

\newcommand{\vim}{\overline{\vi}}
\newcommand{\vjm}{\overline{\vj}}
\newcommand{\vRm}{\overline{\vR}}
\newcommand{\vzm}{\overline{\vz}}
\newcommand{\vphim}{\overline{\vphi}}

\newcommand{\vphiqm}{\overline{v_{\varphi}^2}}
\newcommand{\vRvzm}{\overline{\vR\vz}}
\newcommand{\vRvphim}{\overline{\vR\vphi}}
\newcommand{\vzvphim}{\overline{\vz\vphi}}
\newcommand{\vphivzm}{\overline{\vphi\vz}}
\newcommand{\vivjm}{\overline{\vi\vj}}

\newcommand\sigz{\sigma_z}
\newcommand\sigR{\sigma_R}
\newcommand\sigphi{\sigma_{\varphi}}

\newcommand{\avb}{\langle b\rangle}
\newcommand{\avg}{\langle\gamma\rangle}
\newcommand{\avl}{\langle\lambda^2\rangle}
\newcommand{\avbRi}{\avb^{\rm i}_R}
\newcommand{\avbRo}{\avb^{\rm o}_R}
\newcommand{\avbzo}{\avb^{\rm o}_z}

\newcommand{\sR}{s_R}
\newcommand{\sz}{s_z}

\newcommand{\DMn}{\Delta M_n}
\newcommand{\DN}{\Delta N_n}
\newcommand{\DR}{\Delta R}
\newcommand{\Dz}{\Delta z}
\newcommand{\Nn}{{\cal N}_n} 

\newcommand{\bM}{b_{\rm M}}
\newcommand{\bz}{b_0}

\begin{document} 

   \title{Characterization of the $b$-anisotropy for axisymmetric early-type galaxies in TNG50}

   \author{L. De Deo\inst{1,2}
   \and A. Gebek\inst{3}
   \and M. Baes\inst{3}
   \and L. Ciotti\inst{1}\fnmsep\thanks{luca.ciotti@unibo.it}
   \and S. Pellegrini\inst{1,2}
          }

   \institute{Department of Physics and Astronomy, University of 
            Bologna, Via Gobetti 93/2, I-40129 Bologna, Italy
             \and
             INAF-Osservatorio di Astrofisica e Scienza dello Spazio di Bologna, Via Gobetti 93/3, I-40129 Bologna, Italy
             \and
             Sterrenkundig Observatorium UGent, Krijgslaan 281 S9, 9000 Gent, Belgium
             }

\date{Received, August 12, 2026; accepted, November 17, 2026}

\abstract
 {High-quality observations reveal that internal dynamics of axisymmetric ETGs can be more complicated than commonly adopted in modelizations based on the Jeans equations, with the identity of the radial and vertical velocity dispersions. A well known generalization is obtained by solving the Jeans equations with the prescription of a proportionality factor $b$ between the radial and vertical velocity dispersions; in its simplest form $b$ is assumed constant over the galaxy, while in recent studies $b$ is stratified over planes, with $b=b(z)$.}
{We investigate the properties of the $b$ parameter and of the other kinematical decomposition parameters commonly adopted in modeling in a sample of ETGs taken from the high-resolution cosmological simulations TNG50.}
{From a selection of highly axisymmetric and passive TNG50 (satellite and central) galaxies at equilibrium, we recovered the maps of $b(R,z)$ and of other important kinematic quantities entering the Jeans equations.}
{In all the selected galaxies the only significantly non-zero velocity streaming component is the azimuthal one, and the velocity dispersion tensor is cylindrically aligned. The maps of $b$ are clearly not constant, and not stratified over planes, instead they present a common structure, with the largest values values near the center, and a wedge structure at large (cylindrical) radii. A phenomenological formula for $b(R,z)$ to be used in applications is proposed. Also the other decomposition parameters are non-constant. No significant differences are found as a function of galaxy flattening and mass, while some difference is observed between satellite and central galaxies. }
{The internal dynamics of the TNG50 galaxies present interesting regularities, worth of investigation. On the other hand, in Jeans based studies of real galaxies some simple assumptions for the phenomenological choice of anisotropy and kinematical decomposition parameters do not seem appropriate.}

\titlerunning{Internal dynamics of ETGs in TNG50}
\maketitle

\section{Introduction}

The Jeans equations (hereafter JEs) are a standard tool for the study of the dynamics of galaxies (e.g., see Binney \& Tremaine 2008, hereafter BT08; Bertin 2014; Ciotti 2021, hereafter C21; Cappellari 2008). More general methods of course exist, such as the Schwarzschild (1979) orbit superposition technique (e.g., Cretton et al. 1999, Thomas et al. 2004, Cappellari et al. 2007, van den Bosch et al. 2008), or the ``made-to-measure'' particle models (e.g., Syer and Tremaine 1996, de Lorenzi et al. 2007) but they require a quite sophisticated numerical implementation. From this point of view the use of the more intuitive JEs can provide important information to be used to guide more realistic modeling, and so their study is still of interest, both theoretically and in applications.
In the case of axisymmetric systems, the simplest assumption behind the JEs is that the underlying phase-space distribution function depends on the stellar orbital energy and on the orbital angular momentum component along the symmetry axis\footnote{Cylindrical coordinates $(R,\varphi,z)$ are used, with the $z$-axis aligned with the symmetry axis of the galaxy; a bar over a quantity indicates phase-space average over velocities.}: this implies that the vertical $(\sigz)$ and radial $(\sigR)$ components of the velocity dispersion tensor are identical, and that the velocity dispersion ellipsoid is aligned with the local basis of cylindrical coordinates\footnote{In axisymmetric systems different alignments of the velocity dispersion ellipsoid (for example with the local basis of spherical coordinates, see e.g. Cappellari 2020, 2026), 
can be considered by relaxing the assumption of a two-integral distribution function.} at each point of the system. 

High quality observational data, however, seem to require $\sigz\neq\sigR$, so that a phenomenological, closure relationship between $\sigz$ and $\sigR$ is needed to solve the JEs. A widely adopted choice is the $b$-ansatz successfully used in the Jeans Anisotropic Modeling (JAM) of disk galaxies and Early-Type Galaxies (hereafter ETGs, Cappellari 2008, 2016; e.g., see also Cappellari et al. 2013, Zhu et al. 2016, Loubser et al. 2020, Nitschai et al. 2021, Surti et al. 2024), where the anisotropy parameter $b=\sigR^2/\sigz^2$ is assigned, the JEs are solved, and the solution compared with observational data. 

Albeit phenomenologically satisfactory, the $b$-ansatz (even in its simplest form of a constant $b$) can change the properties of the solution of the JEs with respect to the two-integral case in a non-trivial way. In particular, for some values of $b$, the azimuthal velocity field $\vphiqm$ can become negative, and the model unphysical; time-expensive numerical integrations can be required to exclude unacceptable $b$ values. In order to overcome these problems, a method able to give quantitative information about the kinematical fields produced by the adopted $b$ and assure the positivity of the resulting $\vphiqm$, before solving the JEs, has been developed (De Deo et al. 2024, 2025, hereafter Paper I and Paper II, respectively). The method can be applied not only to constant $b$ but also to the more general plane stratifications, where $b=b(z)$. By using different galaxy models it was shown that in general in order to have a nowhere negative $\vphiqm$, $b$ cannot assume arbitrarily large values. The limits on $b$ depend mainly on the flattening of the  density distribution, and much less on the specific density profile and on the presence of a dark matter halo. These findings are in line with what obtained for real elliptical galaxies (Cappellari 2020), and of realistic mock galaxies modeled through JAM (Wang et al. 2021).

Considering the previous positive aspects, it is important to extend the study of the properties
of the $b$-ansatz. In particular, it should be appreciated that, given an axisymmetric system, with the fields $\sigR(R,z)$ and $\sigz(R,z)$, it is always possible to define aposteriori the positive function $b(R,z)=\sigR^2(R,z)/\sigz(R,z)^2$. So the interesting question here is not if the $b$ ansatz can be applied, but if some characteristic form of $b(R,z)$ is preferred in real systems, e.g. if $b$ is constant, if $b=b(z)$ or even of a more general form, and in case whether this form depends on the galaxy properties. This information of course is of crucial importance not only for proper choices of the $b$ function to be applied for the solution of the JEs, but also for a deeper understanding of the galaxy structure and of how it was established. In this work we study the features of the $b$ function in a sample of axisymmetric ETGs produced in state-of-the-art simulations of galaxy formation and evolution. We consider the TNG50 simulation, which is part of the IllustrisTNG suite of cosmological simulations (Marinacci et al. 2018; Naiman et al. 2018; Nelson et al. 2018; Pillepich et al. 2018b; Springel et al. 2018); its high mass resolution and the large number of ETGs allow for an extended, spatially resolved study of the stellar kinematics. First, we selected a sample of galaxies to be massive, passive, reasonably axisymmetric and at equilibrium, and then we checked whether their kinematics can actually be modeled by using the axisymmetric JEs with the $b$-ansatz, and a cylindrically aligned velocity dispersion ellipsoid. We then investigated the properties of the numerical $b(R,z)$ field, also as a function of the properties of the galaxies and their environment. Finally, we checked whether the observed trend between the upper limit in the amount of $b$-anisotropy and the stellar density flattening holds for these systems as well. The present analysis also offers the interesting opportunity to check how azimuthal motions in the TNG galaxies are divided between ordered rotation and azimuthal velocity dispersion. In the axysimmetric JEs such a decomposition is made aposteriori, because only the introduction of the anisotropy ansatz which is needed to solve the JEs; however also the ordered rotation is relevant especially in modelization works. We notice that examples of Jeans modeling of TNG galaxies exist (e.g., Li et al. 2020; Shi et al. 2024, Pacheco-Arias et al. 2026). 

The paper is organized as follows. In Section 2, we summarize the main results of Papers I and II used in the present study of the selected TNG50 galaxies. In Section 3, the criteria and the methods adopted to construct the sample of axisymmetric TNG50 ETGs, and the procedures used to extract their structural, kinematic and dynamical quantities, are presented. In Section 4 , we present and discuss the results about the properties of the derived $b$-anisotropy, together with those of other important kinematical decomposition parameters of azimuthal motions commonly considered when working with the JEs. Finally, in Section 5, a summary of the main results and some future perspectives are presented.

\section{The Jeans equations}

We summarize here the main results of Papers I and II useful for the present study. We consider the JEs for axisymmetric systems written in cylindrical coordinates
\begin{equation}
\begin{dcases}
    \frac{\partial\rhos\sigz^2}{\partial z}=-\rhos\frac{\partial\Phi}{\partial z}, \\
    \frac{\partial\rhos\sigR^2}{\partial R}-\frac{\rhos\Deltac}{R}=-\rhos\frac{\partial\Phi}{\partial R}, \quad 
    \Deltac \equiv \vphiqm - \sigR^2,
\end{dcases}
\label{eq:3int_jeans}
\end{equation}
where $\rhos(R,z)$ is the stellar density, $\Phi(R,z)$ is the total gravitational potential (that may include the contribution of a DM halo, a central black hole, etc.), $\sigz$ and $\sigR$ are respectively the vertical and radial velocity dispersions, and $\vphiqm = \vphim^2 + \sigphi^2$, where $\vphim$ is the azimuthal streaming velocity, and $\sigphi^2$ is the azimuthal velocity dispersion. As in the two-integrals case (when $\sigR=\sigz$), it is assumed that $\vRm=\vzm =0$, and that the off-diagonal terms of the velocity dispersion tensor vanish (i.e., the velocity dispersion tensor is aligned with the local basis of cylindrical coordinates).

The solution of the first JE with $\rhos\sigz^2 =0$ for $z\to\infty$ reads
\begin{equation}
    \rhos\sigz^2 = \int_z^\infty \rhos\frac{\partial\Phi}{\partial z'}dz'.
\label{eq:2int_JE_sol}
\end{equation}
The radial JE contains the two unknown functions $\vphiqm$ and $\sigR^2$, and a closure is needed for its solution. In the $b$-ansatz, a value of $b$ is fixed and 
\begin{equation}
    \sigR^2 = b \sigz^2,
    \label{eq:b_ans_def}
\end{equation}
where $b=1$ corresponds to the two-integrals case. Following Paper I, under the assumption that $b$ is independent of $R$, i.e., that at most $b=b(z)$, from the the radial JE
\begin{equation}
\Deltac = b\frac{R D}{\rhos} + R\frac{\partial\Phi}{\partial R},\quad D \equiv \frac{\partial\rhos\sigz^2}{\partial R},
\label{eq:b_ans_deltaR}
\end{equation}
and once $\Deltac$ is known
\begin{equation}
\vphiqm = \Deltac + \sigR^2 = bB + R\frac{\partial\Phi}{\partial R},\quad B \equiv \frac{R D}{\rhos} + \sigz^2.
\label{eq:b_ans_vphiqm}
\end{equation}
Notice that the $B$ and $D$ functions are independent of the specific $b$ adopted, and for a given model can present positive and negative regions over the $(R,z)$ plane: only the choices of $b$ leading to $\vphiqm\ge 0$ everywhere are physically acceptable.  As discussed in Papers I and II, in the simplest case of a constant $b$, one can determine the values $\bM$ and $\bz$ such that $\vphiqm\leq 0$ somewhere for $b\geq\bM$, and $\Deltac\leq 0$ somewhere for $b\geq\bz$.

Models with $\Deltac\ge 0$ allow for a decomposition of $\vphiqm$ in its ordered and dispersion components as
\begin{equation}
    \vphim = k \sqrt{\Deltac}, \; \; \sigphi^2 = \sigR^2 + (1-k^2)\Deltac,
    \label{eq:satoh_dec_bans}
\end{equation}
analogous of the widely used Satoh (1980) decomposition for two-integrals systems,
where $\Delta=\vphiqm - \sigz^2\geq 0$ is used instead of $\Deltac$.
 $k^2<1$ corresponds to tangential anisotropy ($\sigphi>\sigR$), while radial anisotropy ($\sigphi< \sigR$) is obtained for $k^2>1$; $k^2=1$ and $b=1$ describe complete isotropy. Notice that $k$ needs not to be constant nor positive, but it can be a function of $(R,z)$, provided that $\sigphi^2\geq 0$ (e.g., see Ciotti \& Pellegrini 1996). A different decomposition, that can be applied also to models with $\Deltac <0$, is obtained by choosing the parameter
\begin{equation}
    \gamma\equiv 1 - \frac{\sigphi^2}{\sigR^2} \implies \vphim^2 = \Deltac + \gamma\sigR^2,
    \label{eq:gamma_def}
\end{equation}
(Cappellari et al. 2007). The positivity of $\vphim^2$ and $\sigphi^2$ imposes that 
\begin{equation}
    -\frac{\Deltac}{\sigR^2} \le \gamma\le 1,
    \label{eq:gamma_cons}
\end{equation}
and in general $\gamma$ can depend on $R$ and $z$. From equation (\ref{eq:gamma_def}), a negative (positive) $\gamma$ corresponds to tangential (radial) anisotropy; $\gamma = 0$ and $b=1$ describe complete isotropy. Notice that, from equation (\ref{eq:gamma_cons}), $\gamma$ can be either positive or negative if $\Deltac > 0$; instead, if $\Deltac < 0$ necessarily 
$\gamma>0$, and  the velocity dispersion tensor is radially anisotropic. Finally, we can consider the decomposition 
\begin{equation}
\vphim = \lambda \sqrt{\vphiqm},\quad\sigphi^2 = (1-\lambda^2)\vphiqm,\quad 
0\leq\lambda^2\leq 1,
    \label{eq:lambda_def}
\end{equation}
(e.g. Caravita et al. 2021, PaperII), that, as the previous one\footnote{The decompositions in equation (\ref{eq:gamma_def})-(\ref{eq:lambda_def}) are not independent, as for given $\vphiqm$ and $\sigR^2$
\begin{equation}
(1-\lambda^2)\vphiqm = (1-\gamma)\sigR^2, 
\quad {\rm i.e.}\quad 
(1-\lambda^2)\Deltac=(\lambda^2-\gamma)\sigR^2.
\end{equation}},
can be adopted also for models with $\Deltac \leq 0$. In the most general case $\lambda$ can be a function of position inside the galaxy; the equality $\sigR=\sigphi$ is realized for $\lambda^2 =\Deltac/(\Deltac +\sigR^2)$.

In the following we focus on a sample of ETGs produced by high-resolution TNG50 cosmological simulations to test whether, and in case how well, the massive and passive galaxies (restricting to those with axisymmetric stellar distributions) satisfy the JEs in equation (\ref{eq:3int_jeans}); if this is the case, we investigate what are the properties of the $b$ function (if constant, plane stratified or more general; if different from galaxy-to-galaxy or presenting some common features), and of the decomposition parameters $k$, $\gamma$, and $\lambda$.

\section{The TNG50 simulation}

This Section describes the selection procedures used to construct the galaxy sample, and how the structural and kinematic quantities of the selected galaxies are determined.

\subsection{Sample and selection criteria}
\label{sec:tng50}

The IllustrisTNG simulations are a suite of state-of-the-art cosmological simulations of galaxy formation and evolution (Marinacci et al. 2018; Naiman et al. 2018; Nelson et al. 2018; Pillepich et al. 2018b; Springel et al. 2018). Compared to the previous Illustris simulations (see e.g. Vogelsberger et al. 2014), they contain a new implementation of black hole feedback (Weinberger et al. 2017), as well as updates to the galactic wind feedback, stellar evolution and gas chemical enrichment models (Pillepich et al. 2018a). For example, the new AGN feedback model is responsible for the quenching of galaxies in massive halos and for the production of red and passive galaxies at late times, alleviating the discrepancies with observational results at the massive end of the halo mass function (Weinberger et al. 2017; Nelson et al. 2018; Donnari et al. 2019). The faster stellar winds in TNG reduce the star formation at all masses and all times, resulting in a suppressed $z = 0$ galaxy stellar mass function for $\Mstar\lesssim 10^{10}\Msun$, and smaller galaxy sizes. Overall, the TNG model has been demonstrated to agree satisfactorily with many observational constraints (see e.g. Genel et al. 2018; Nelson et al. 2018) and to return a reasonable mix of morphological galaxy types (Rodriguez-Gomez et al. 2019). 

For the present dynamical analysis, we consider TNG50-1, the main high-resolution IllustrisTNG50 run including the full TNG physics model. In particular, the computational box size is $\simeq 50$ Mpc$^3$ (comoving), and the mass resolution $\simeq 8.5 \times 10^4\Msun$ is typical of zoom-in simulations (Pillepich et al. 2019; Nelson et al. 2019); the adopted cosmological parameters ($\Omega_{\rm m}$ = 0.3089, $\Omega_{\rm b}$ = 0.0486, $\Omega_{\Lambda}$ = 0.6911 and $H_0$ = 67.74 km s$^{-1}$ Mpc$^{-1}$) are based on the latest Planck results (Planck Collaboration et al. 2016). Overall, the main advantage of TNG50 for the present study is the high spatial and mass resolution, even though the limited simulation volume results in a relatively small final sample.

As a first step, from the TNG50 cutout at redshift $z=0$, we selected the galaxies with stellar mass in the range $10^{10.5} \le\Mstar/\Msun\le 10^{12}$, and with a low specific star formation rate sSFR $\le 10^{-12}\, {\rm yr}^{-1}$, obtaining a sample of 81 bona-fide ETGs.

\subsection{Structural analysis}
\label{sec:structure}

In the second step of the analysis we restricted the sample by retaining only those ETGs with a clear axisymmetric structure. In practice for each of the 81 galaxies we computed the second-order mass tensor
\begin{equation}
 \mathcal{S}_{ij} = \frac{\sum_k\mk\, r_{ki}\, r_{kj}}{\sum_k\mk},
 \label{eq:shape_tens}
 \end{equation}
where $r_{ki}$ is the $i$-component of the position vector of the $k$-th stellar particle, and the summation is over the particles within $2\rh$, where $\rh$ is the stellar, volumetric half-mass radius (e.g., see Zemp et al. 2011, Diemer et al. 2018). As well known for an ellipsoidally stratified density distribution of total mass $M$ and semiaxes $a\geq b\geq c$ the eigenvectors of $\mathcal{S}$ give the direction of the ellipsoid principal axes, while the eigenvalues can be written as $M\, h\, a^2/3$, 
$M\, h\, b^2/3$, and $M\, h\, c^2/3$, where the shape factor $h$ depends on the specific density profile but is independent of $a$, $b$, and $c$ (e.g., see exercises 2.17 and 2.18 in C21). Accordingly, we diagonalized ${\cal S}$ of each galaxy,  defined the axial ratios as the square root of the ratio of the eigenvalues, and selected only the systems with an intermediate-to-long axial ratio $b/a \ge 0.95$; the galaxy flattening was defined as $q=c/a$. 

\begin{figure}[ht!]
    \centering
    \includegraphics[width=0.9\linewidth]{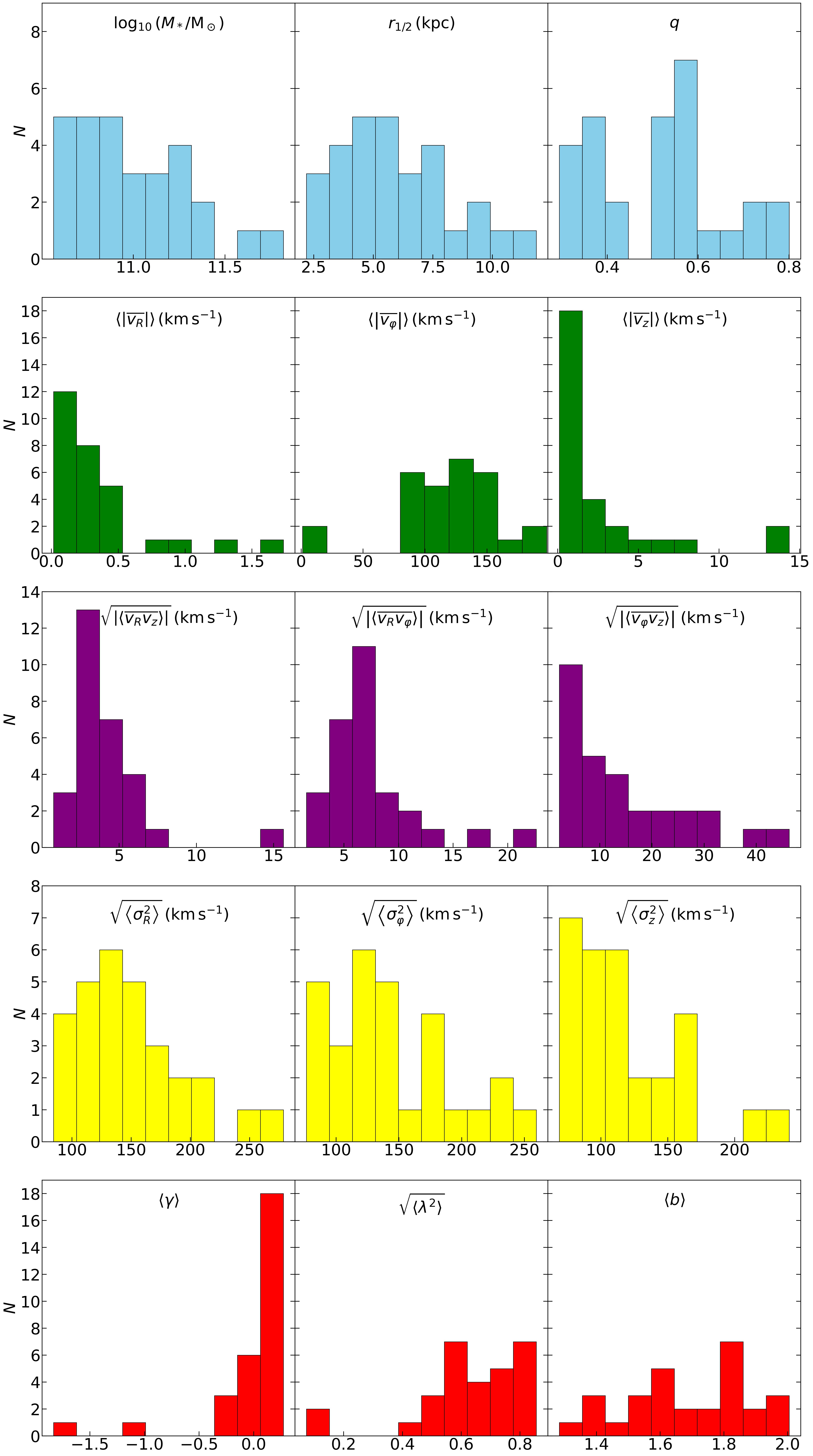}
    \caption{Number distribution of the 29 galaxies in Table \ref{table:pmore} as a function of their stellar mass $\Mstar$, half-mass radius $\rh$, axial ratio $q$ (first row), of the absolute value of the mass-weighted streaming velocity components $\langle\vRm\rangle$, $\langle\vphim\rangle$, $\langle\vzm\rangle$ (second row), of the square root of the absolute value of the mass-weighted off-diagonal terms $\langle\vRvzm\rangle$, $\langle\vRvphim\rangle$, $\langle\vphivzm\rangle$  of the second-order velocity tensor (third row), of the mass-weighted velocity dispersions $\langle\sigR^2\rangle$, $\langle\sigphi^2\rangle$, $\langle\sigz^2\rangle$ (fourth row), and finally of the mass-weighted azimuthal velocity decomposition parameters $\avg$, $\sqrt{\avl}$, and of the mass-weighted anisotropy parameter $\avb$ (fifth row). Mass weighted quantities are computed according to equation (\ref{eq:mass_average}).}
    \label{fig:figure1}
\end{figure}

Table \ref{table:pmore} summarizes the main structural properties of the final sample of 29 galaxies, divided in the two groups of 15 Satellites and 14 Central galaxies. They are listed in each group for increasing mass; the three top panels of Figure \ref{fig:figure1} show the number distribution of the 29 galaxies as a function of $\Mstar$, $\rh$, and $q$; $\rh$ and $q$ reassuringly span the range of values commonly shown by real ETGs. We stress that the lower threshold 0.95 for axysimmetry is quite strict, and reflects a compromise between selecting systems that are as close to axisymmetric as possible while retaining a sufficiently large sample of galaxies.

For each galaxy we then introduced a new Cartesian reference system aligned with the eigenvectors of ${\cal S}$, and with the $z$-axis along the short (symmetry) axis. In the new reference system, we also introduced standard cylindrical coordinates $\varphi$ and $R = \sqrt{x^2 + y^2}$. The galaxies in the final sample are highly axisymmetric, however in order to reduce the numerical noise due to the finite number of particles we considered the azimuthally averaged density 
\begin{equation}
    \overline{\rhos}(R,z)\equiv\frac{1}{2\pi}\int_0^{2\pi}\rhos(R,\varphi,z)\,d\varphi.
\end{equation}
Being the numerical $\rhos(R,\varphi,z)$ discrete, the region $R\leq  2\rh$ and $\vert z\vert\leq\rh$ was first subdivided into a regular grid of ``pixels'' of sides $\DR =\Dz$ and center $(R,z)$, with vertices $R_{\pm}=R\pm\DR/2$ and $z_{\pm}=z\pm\Dz/2$. Each pixel is the vertical section of a ``torus'' produced by a $2\pi$ rotation of the pixel around the $z$-axis, so that the smoothed density of each torus is
\begin{equation}
    \overline{\rhos}(R,z) 
    = \frac{\DMn}{2\pi R\DR\Dz}= {\Nn\over 2\pi R}{\DMn\over\DN},
 \label{eq:rhopixel}
\end{equation}
where $\DMn=\sum_k\mk$ is the mass of the $n$-th torus, $\mk$ is the mass of a stellar particle, and the sum extends over the $\DN$ particles of the torus. Finally,
\begin{equation}
\Nn = {\DN\over\DR\Dz}, 
\label{eq:npixel}
\end{equation}
would be the pixel surface density if all the $\DN$ particles are imagined on the pixel. For simplicity of notation, from now on we set $\rhos(R,z)=\overline{\rhos}(R,z)$. The size of the pixels was fixed to guarantee $\DN\gtrsim 200$ in each torus in the main body of the selected galaxies, a condition that allow a robust determination of the kinematic fields (Harborne et al. 2024). We found that an optimal compromise between the request on $\DN$, and spatial resolution is obtained for $\DR =\Dz = 200$ pc; we also considered the case of $\DR =\Dz = 100$ pc, but then the recommended critical $\DN$ was only reached in the very central region of the galaxies. 

\subsection{Kinematical and dynamical analysis}
\label{sec:kinematics}

The kinematic fields of the galaxies were determined following the same method used to compute $\rhos$. In the eigenvectors reference system, for each stellar particle of velocity $(\vx,\vy,\vz)$
\begin{equation}
\vR=\vx\cos\varphi + \vy\sin\varphi,\quad
\vphi= -\vx\sin\varphi + \vy\cos\varphi.
\end{equation}
Then for each torus
\begin{equation}
\vim= \frac{\sum_k\mk\vki}{\DMn},\quad
\vivjm= \frac{\sum_k\mk\vki\vkj}{\DMn},\quad
\sigma_{ij}^2\equiv\vivjm -\vim\, \vjm,
\label{eq:fmeantori}
\end{equation}
where the index $i$ indicates $R,\, \varphi,\, z$; in particular $\sigma_i^2 = \overline{\vi^2} - \vim^2$. Once the fields are known, the functions $b(R,z)$ and the decomposition $k(R,z)$, $\gamma(R,z)$ and $\lambda(R,z)$ for the TNG50 ETGs were determined. Finally, for a generic dynamical property $f$, we computed its mass-weighted average as 
\begin{equation}
    \langle f\rangle\equiv \frac{\sum_k\mk f_k}{\Delta M (2\rh)} = \frac{\sum_n \DMn \overline {f_n}}{\Delta M (2\rh)},
\label{eq:mass_average}
\end{equation}
where the first sum extends to the particles inside the sphere of radius $2\rh$, while the second sum extends over the tori contained in the sphere, with $\overline {f_n}$ of each torus defined as in equation (\ref{eq:fmeantori}); the second expression was used to compute the mass-weighted values of $\avg$, $\avl$, and $\avb$.

\begin{figure}[ht!]
        \centering
        \includegraphics[width=0.9\linewidth]{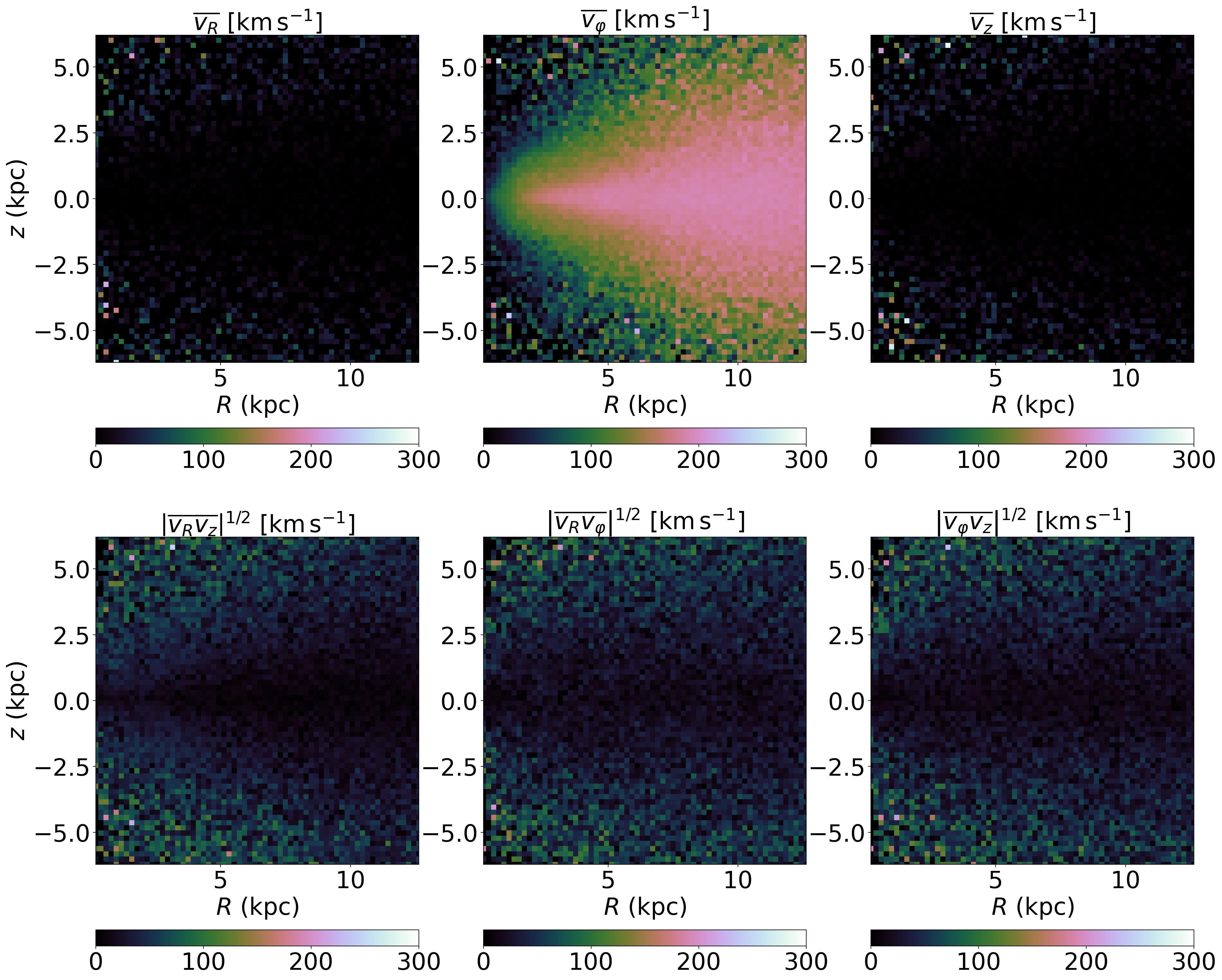}
        \includegraphics[width=0.9\linewidth]{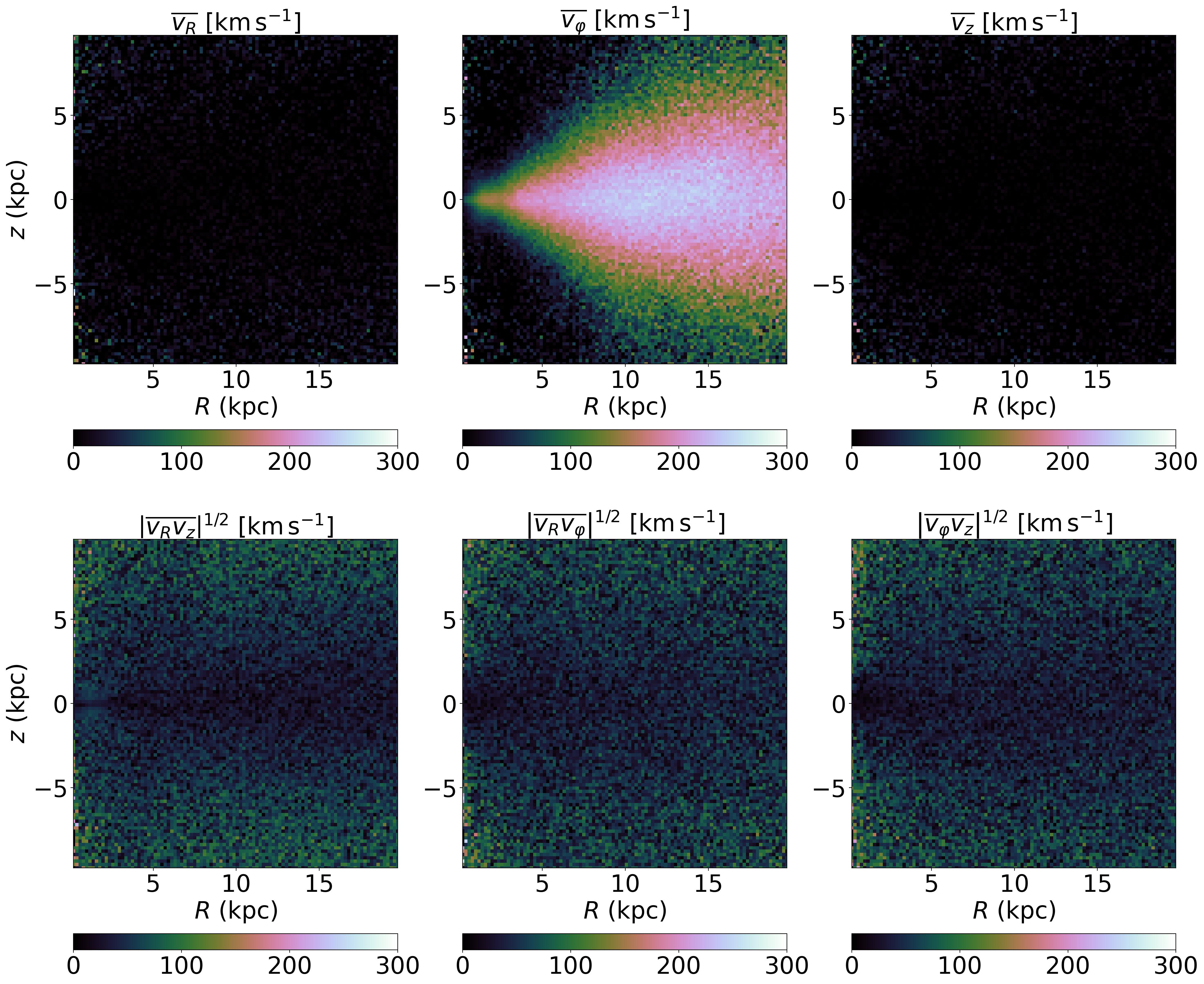}
\caption{Maps on the $(R,z)$ plane ($R \leq2\rh$,  $\vert z\vert\leq\rh$) of the streaming velocities (first row) and of the off-diagonal components of the velocity dispersion tensor (second row) for the satellite galaxy TNG000010 ($\Mstar= 6.12 \times 10^{10}\Msun$, $\rh = 6.41$ kpc, $q\simeq 0.3$), and for the more massive and rounder central galaxy TNG390932 ($\Mstar = 2.62 \times 10^{11}\Msun$, $\rh = 9.97$ kpc, $q\simeq 0.6$, third and fourth row). Maps of $\sigz$, $\sigR$, and $\sigphi$ (with the same color scale) of the two galaxies are shown respectively in the fourth panel of the first row, and in the third panel of the last row in Figures \ref{fig:figure3}-\ref{fig:figure4}-\ref{fig:figure5}.}         
\label{fig:figure2}
\end{figure}
  
Finally, in order to check how well the selected 29 TNG50 galaxies are described by the time-independent JEs (\ref{eq:3int_jeans}) we constructed the left and right hand sides of the vertical and radial JEs directly from the numerical simulations, and examined how much they differ. A necessary ingredient for this test is the knowledge of the galaxy gravitational potential (stars plus DM). Accordingly, for each galaxy we computed  $\Phi(R,z)$ as the angle-average of
\begin{equation}
    \Phi = -G\sum_k \frac{\mk}{\sqrt{(x - \xk)^2+ (y - \yk)^2 + (z - \zk)^2 + \epsilon^2}},
    \label{eq:potential}
\end{equation}
where the sum goes over all the stellar and DM particles in the subhalo hosting the galaxy, and the softening length $\epsilon$ is set equal to the pixel scale $\DR=\Dz$. 

\section{Results} 
\label{sec:results}

 A first point of interest is how well the 29 TNG50 ETGs, selected only by axisymmetry and negligible specific star formation rate at the present epoch, are described by the time-independent JEs in equation (\ref{eq:3int_jeans}). The second row of Figure \ref{fig:figure1} shows the number distribution of the absolute value of the mass-weighted streaming velocity fields $\langle\vRm\rangle$, $\langle\vphim\rangle$, and $\langle\vzm\rangle$: the values of $\langle\vRm\rangle$ and $\langle\vzm\rangle$ are between $\approx$ one and two orders of magnitude smaller than those of $\langle\vphim\rangle$. Similarly, the values of the mass-weighted off-diagonal terms $\langle\vivjm\rangle$ (third row) are $\approx$ one order of magnitude smaller than those of the three principal velocity dispersions (fourth row), a robust indication of the alignment of the velocity dispersion tensor with cylindrical coordinates. These findings are exemplified by Figure \ref{fig:figure2}, where we show the $(R,z)$ maps of the streaming velocities, and of the off-diagonal components of the velocity dispersion tensor for the satellite galaxy TNG000010, and for the more massive and rounder central galaxy TNG390932; the velocity dispersion fields $\sigz$, $\sigR$, and $\sigphi$ of the two galaxies are shown (with the same color scale) in Figures \ref{fig:figure3}, \ref{fig:figure4}, and \ref{fig:figure5}. As an additional test of the overall equilibrium configuration of the selected ETGs, from their simulation data we constructed numerically the l.h.s. and the r.h.s. of the vertical and radial JEs. We found that over the region of the $(R,z)$ plane well sampled according to the criterion discussed in Section 3.2, the two sides of the JEs differ less than $\approx 20\%$. This represents a reassuring result, also considering the use of numerical differentiation of numerically derived quantities, that supports the assumption of global equilibrium of the studied galaxies.

We can now describe the main properties of the kinematical fields of the selected ETGs. In general, the streaming velocities and the off-diagonal terms of the velocity dispersion tensor, are well represented by those of the two galaxies in Figure \ref{fig:figure2}, confirming global equilibrium and cylindrical alignment, and will not be discussed further. The maps of the velocity dispersions $\sigz$, $\sigR$, and $\sigphi$ are instead shown Figures \ref{fig:figure3}, \ref{fig:figure4}, and \ref{fig:figure5} for all the galaxies in Table \ref{table:pmore}; blue and red top labels over each map indicate respectively satellite and central galaxies. In each of the two groups, galaxies are ordered by increasing stellar mass, as also evidenced by the values of the velocities, with colors changing systematically from dark to light, moving from low mass to high mass systems. The black lines in each box show the stellar isodensity contours; notice how a significant number of cases, both for satellites and centrals, presents an evident disky structure. 

In general $\sigz$ (Figure \ref{fig:figure3}) tends to be lower in the equatorial plane at large radii, and higher near the center and above and below the equatorial plane, producing a quite characteristic hourglass configuration, more evident in flatter systems; this feature had already been evidenced as a common property in different analytical models (e.g., see Negri et al. 2014, Ciotti et al. 2021). The maps of $\sigR$ (Figure \ref{fig:figure4}) are qualitatively similar to those of $\sigz$; however for several galaxies $\sigR > \sigz$ over vast regions of the $(R,z)$ plane, a first clear indication that important $b$-anisotropy is present.  The maps of $\sigphi$ (Figure \ref{fig:figure5}) show instead a variety of cases when compared to those of $\sigR$. Even if their general aspect is quite similar, with $\sigphi$ near the equatorial plane decreasing for increasing $R$ while remaining larger than $\sigR$, for some galaxies $\sigphi$ increases at increasing $R$, and/or shows interesting structures (see in particular the satellites TNG264885 in the second row, TNG143883 and TNG000001 in the third row, and the central galaxies TNG543114 and TNG450916 in the fifth row, and TNG390932 in the last row). We will return on this point when discussing the parameter $\gamma(R,z)$.

About $\Deltac(R,z)$ and $\Delta(R,z)$, perhaps the most interesting result is the evidence of large regions with negative values, showing that the internal dynamics of the selected TNG50 ETGs cannot be described by Satoh decomposition or its variants. Figure \ref{fig:figure6} shows the field $\Deltac(R,z)$, with the white regions corresponding to negative values; the white lines are lines of constant ${\cal N}$ corresponding to 200 particles-per-pixel (the innermost), and declining outside to 150, 100, and 50, so that we consider the existence of negative values of $\Deltac(R,z)$ a robust result. Overall the shape of constant $\Deltac$ contours remind quite closely that of $\Deltac$ for the analytical models discussed in Paper I and II. The fields of $\Delta$ are quite similar to those of $\Deltac$, with large negative regions, and so they are not shown. Notice how the shape of the white contours is not the same as that of the black isodensity contours in Figures \ref{fig:figure3}-\ref{fig:figure4}-\ref{fig:figure5}; this is expected from the geometrical scaling between ${\cal N}$ and $\rhos$ (see equation [\ref{eq:rhopixel}]), which produces the characteristic closure of the white lines near the $z$-axis.

In order to obtain a hint of how the kinematical fields would appear to an observer, in Figs. \ref{fig:figure7}-\ref{fig:figure8} we show the line-of-sight projection of the streaming ($v_{\rm los}$) and velocity dispersion ($\sigma_{\rm los}$) fields, computed following equations (20)-(23) of Caravita et al. (2021); due to the highly symmetric structure of the selected models, we arbitrarily projected them along their $y$-axis. The resulting $v_{\rm los}(x,z)$ fields reveal a common (more or less marked) wedge structure, as expected from the analogous structure of the internal streaming velocity field $\vphim(R,z)$, shown for two representative galaxies in Figure \ref{fig:figure2}. Also $\sigma_{\rm los}(x,z)$ reflects quite closely the main features of the intrinsic velocity dispersion fields; at this stage we do not pursue further the study of ``observational'' properties of the models, that could be extended to include for example the bidimensional maps of line profile parameters such as $h_3$ and $h_4$.

Figures \ref{fig:figure9}-\ref{fig:figure10} show the maps of the decomposition parameters $\gamma (R,z)$ and $\lambda^2(R,z)$, following the same ordering in the previous Figures \ref{fig:figure3}-\ref{fig:figure6}; $\gamma$ and $\lambda$ are both clearly not constant over the galaxy body. In some cases $\gamma$ tends to be negative (blue-green color, corresponding to $\sigphi\geq\sigR$) in a wedge-shaped region symmetric with respect to the equatorial plane, as particularly evident for TNG143883 in the third row of Figure \ref{fig:figure9}. In other cases, the negative region is even larger (e.g., for TNG390932, in the last row). Instead $\gamma$ is positive (red-orange color, corresponding to $\sigphi\leq\sigR$) in the central regions of the galaxies, in some galaxy over the whole equatorial plane (TNG572328 in the fourth row), and in some cases $\gamma$ is even increasing for increasing $R$ (TNG000005 in the second row).
Figure \ref{fig:figure10} shows that the decomposition parameter $\lambda^2$ is generally near its maximum value of 1 in a region including the equatorial plane, naturally reminescent of the region where $\vphim$ is large, as shown by the two galaxies in Figure \ref{fig:figure2}; however significant differences are present from galaxy to galaxy. The mass-weighted values $\avg$ and $\langle\lambda^2\rangle$ of the two decomposition parameters are reported in Table \ref{table:pmore} and also shown in the last row of Figure \ref{fig:figure1}. Interestingly, Table \ref{table:pmore} indicates that negative values of $\avg$ appear preferentially in Central galaxies rather than in Satellites.

In Figure \ref{fig:figure11} we finally show $b(R,z)$ for the 29 selected ETGs.
As for the decomposition parameters $\gamma$ and $\lambda$, also $b$ is not constant, with values larger than unity in the equatorial plane, and even larger than 2 near the center. These values are attained well inside the regions with an optimal number of particles per pixel. $b$ tends to decrease at increasing $\vert z\vert$, and for many galaxies the lowest values are contained in the hourglass region around the symmetry axis characteristic of $\sigR$ and $\sigz$. However, also along the equatorial plane, $b$ is not constant: in some cases $b$ steadily decreases for increasing $R$ but in many others, after a first decrease, it increases again, in a characteristic triangular region extending to larger $\vert z\vert$ for increasing $R$ (e.g., see the satellite galaxies TNG000009 in the first row, TNG000005, TNG096763, TNG096764 in the second row, TNG000002 in the third row). The last panel of Figure \ref{fig:figure1} shows the number distribution of the mass-weighted values $\avb$, where all values are unsurprisingly larger than unity, a consequence of the spatial distribution of the large $b(R,z)$ values that are located in the high-density equatorial regions.

\begin{figure}[ht!]
            \centering
            \includegraphics[width=0.9\linewidth]{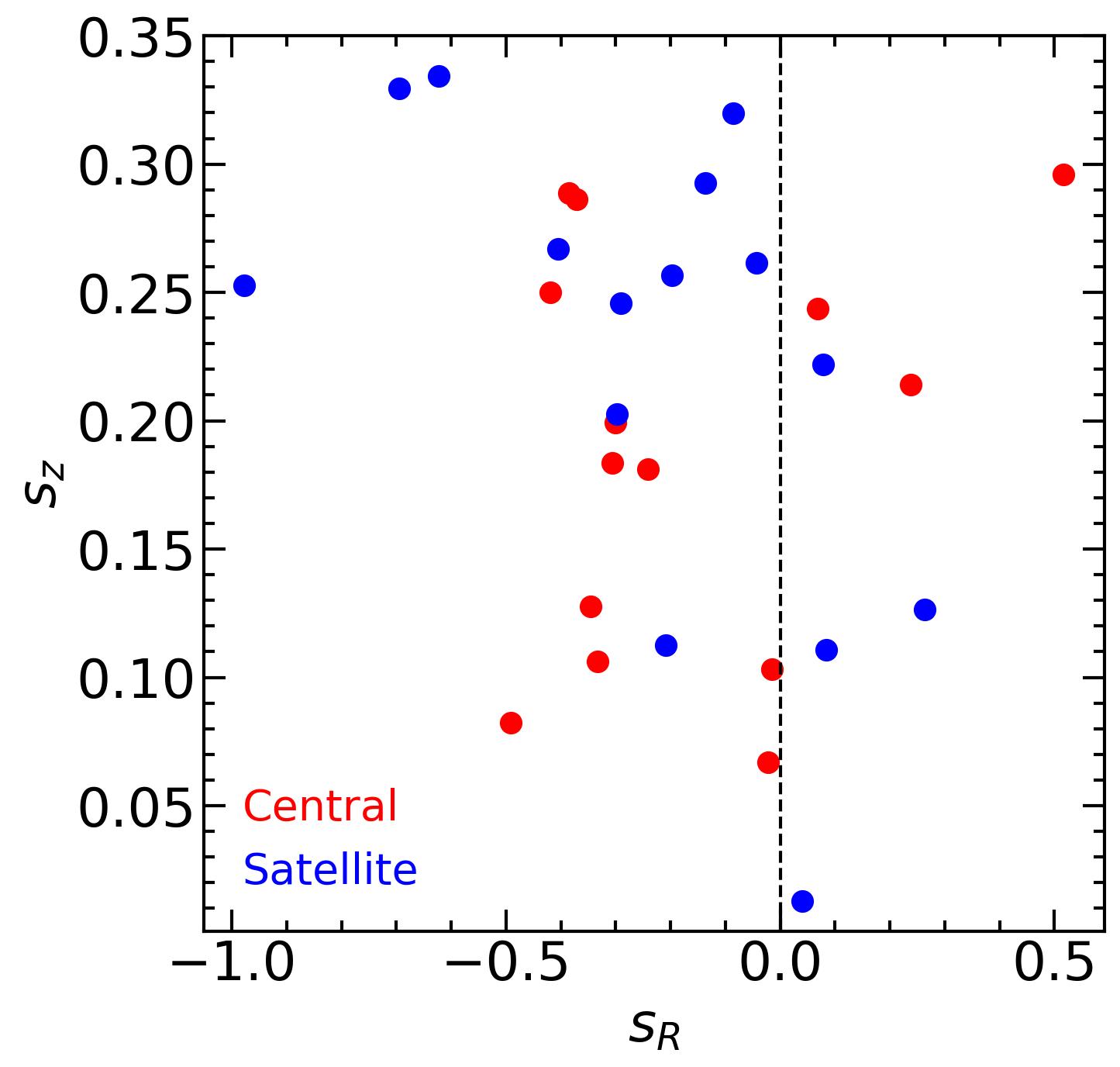}
            \caption{The $\sR$ and $\sz$ parameters (equation (\ref{eq:s_par})) for the $29$ TNG50 ETGs in Table \ref{table:pmore}, with Satellites represented by blue points, and Central galaxies by red points.}
            \label{fig:figure12}
        \end{figure}

These results indicate that not only the choice of a constant $b$ is excluded for the TNG50 ETGs, but also that its generalization as $b(z)$ studied in Paper I and II is still too idealized, especially at large $R$. To better quantify the variation of $b$, for each galaxy we constructed the two parameters
\begin{equation}
\sR \equiv 1 - \frac{\avbRo}{\avbRi},\quad 
\sz \equiv 1 - \frac{\avbzo}{\avbRo},
\label{eq:s_par}
\end{equation}
where $\avbRi$ and $\avbRo$ are the mass-weighted values of $b$ obtained by restricting to an internal region of $0.25\rh\leq R\leq 0.75\rh$ of the equatorial plane, and to an outer one with $1.25\rh\leq R\leq 1.75\rh$. Therefore, $\sR$ measures how fast the $b$ parameter is decreasing ($\sR >0$) or increasing ($\sR<0$) along the equatorial plane. With $\avbzo$ we instead measure how much $b$ changes moving away from the equatorial plane: due to the particular wedge shape of $b$ in the outer regions, we consider the mass-weighted values of $b$ in the two symmetric regions above and below the equatorial plane with $1.25\rh\leq R\leq 1.75\rh$ and $\vert z\vert=0.5\rh$; $\avbzo$ is taken to be the arithmetic mean of the two values. Positive values of $\sz$ therefore indicate a decrease of $b$ at increasing $\vert z\vert$, while $\sz\lesssim 0$ indicates a $b$ function constant with $\vert z\vert$ or even increasing. Figure \ref{fig:figure12} shows the pairs $(\sR, \sz)$ for the $29$ galaxies (blu points for Satellites, and red for Central galaxies): the values for each galaxy are reported in Table \ref{table:pmore}. It is apparent how for all galaxies $\sz > 0$, with values ranging from $\simeq 0$ to a maximum of $\simeq 0.35$. Instead,  positive and negative values of $\sR$ appears, with a clear predominance of negative values, i.e., $b$ tend to increase for in increasing $R$, in agreement with the visual inspection of Figure \ref{fig:figure11}. Interestingly, if we consider the median ($\simeq 0.22$) of the $\sz$ distribution, the upper half of the diagram is dominated by Satellite galaxies, while no difference of the two types is revealed by the $\sR$ distribution. We checked possible systematic dependencies of $\sR$ and $\sz$ on other galaxy properties, such as mass, flattening, size, richness, but all these quantities showed a significantly lower (or negligible) effect.

\begin{figure}[ht!]
       \centering
        \includegraphics[width=0.97\linewidth]{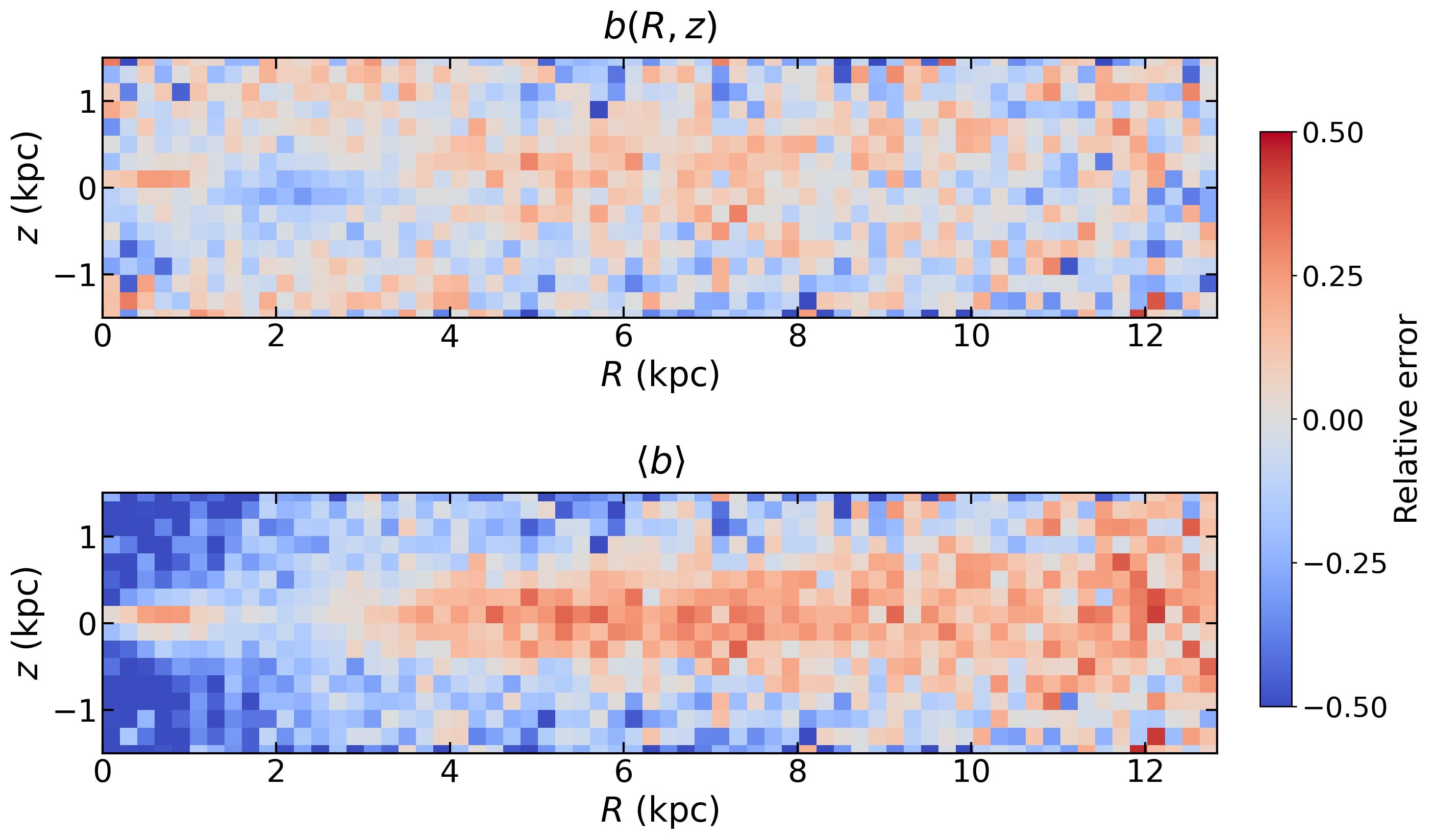}
        \caption{Relative error $1-b_{\rm fit}/b_{\rm num}$ for the TNG000010 galaxy, over the numerically well-sampled region (see Section 3.2). In the top panel $b_{\rm fit}$ is the best fit of the numerical $b(R,z)$ in Figure \ref{fig:figure11}, obtained from equation (\ref{eq:brzfit}) with $c=0$, $E_0\simeq 4.75$, $E_1\simeq -0.0056$, $E_2\simeq 0.0001$, $E_3\simeq 0.002$, $n\simeq 0.166$, $a\simeq 5.18$, $p\simeq 0.22$, and with lenghts in kpc units. In the bottom panel $b_{\rm fit}$ is the spatially constant mass-weighted $\avb$ given in Table \ref{table:pmore}.}
        \label{fig:figure13}
    \end{figure}

The main properties of the $b(R,z)$ fields can be phenomenologically reproduced by the mathematically motivated function
\vskip 0.01truemm
\begin{equation}
b(R,z)={E(R)\over\sqrt[n]{1+[\,\vert z\vert/W(R)\,]^n}} + c,\quad n>0,\; c\geq 0,
\label{eq:brzfit}
\end{equation}
so that on the equatorial plane $b(R,0)=E(R)+c$. Also, notice that $b(R,z)\sim b(R,0)$ almost independent of $z$ in the region $\vert z\vert\leq W(R)$, and $b(R,z)\sim E(R)W(R)/\vert z\vert + c $ for $\vert z\vert\gg W(R)$, so that for suitable choices of the functions $E(R)$ and $W(R)$ it is possible to reproduce the regions of high and low $b$ values observed in Figure \ref{fig:figure11}. From inspection, it appears quite natural to use functions such as
$W(R)=(a + R)^{1/p}$, with $a$ and $p$ positive numbers, and  
\begin{equation}
E(R)={E_0 + E_1 R^2 + E_2 R^4 \over (1+ E_3 R^2)^2},
\end{equation}
where the values of the $E_i$ constants can be chosen to fix the value of $b$ 1) at the center, 2) at large radii, and 3) to produce a minimum at a given position on the equatorial plane (as observed for example in TNG454171 or TNG482155). As preliminary tests, we fitted the $b(R,z)$ fields of some galaxy, obtaining percentual errors between the fits and the numerical data of $\lesssim 20\%$ over the regions well sampled by the numerical simulations. For illustration in Figure \ref{fig:figure13} we show the relative error produced by the fitted $b(R,z)$ for the galaxy TNG000010 (top panel), to be compared with the error obtained by assuming the mass-weighted constant value $\avb$ in Table \ref{table:pmore} (bottom panel): of course, even better agreement could be obtained by tailoring the $W(R)$ formula, but at this stage we do not explore this possibility any further.

\begin{figure}[ht!]
       \centering
        \includegraphics[width=0.825\linewidth]{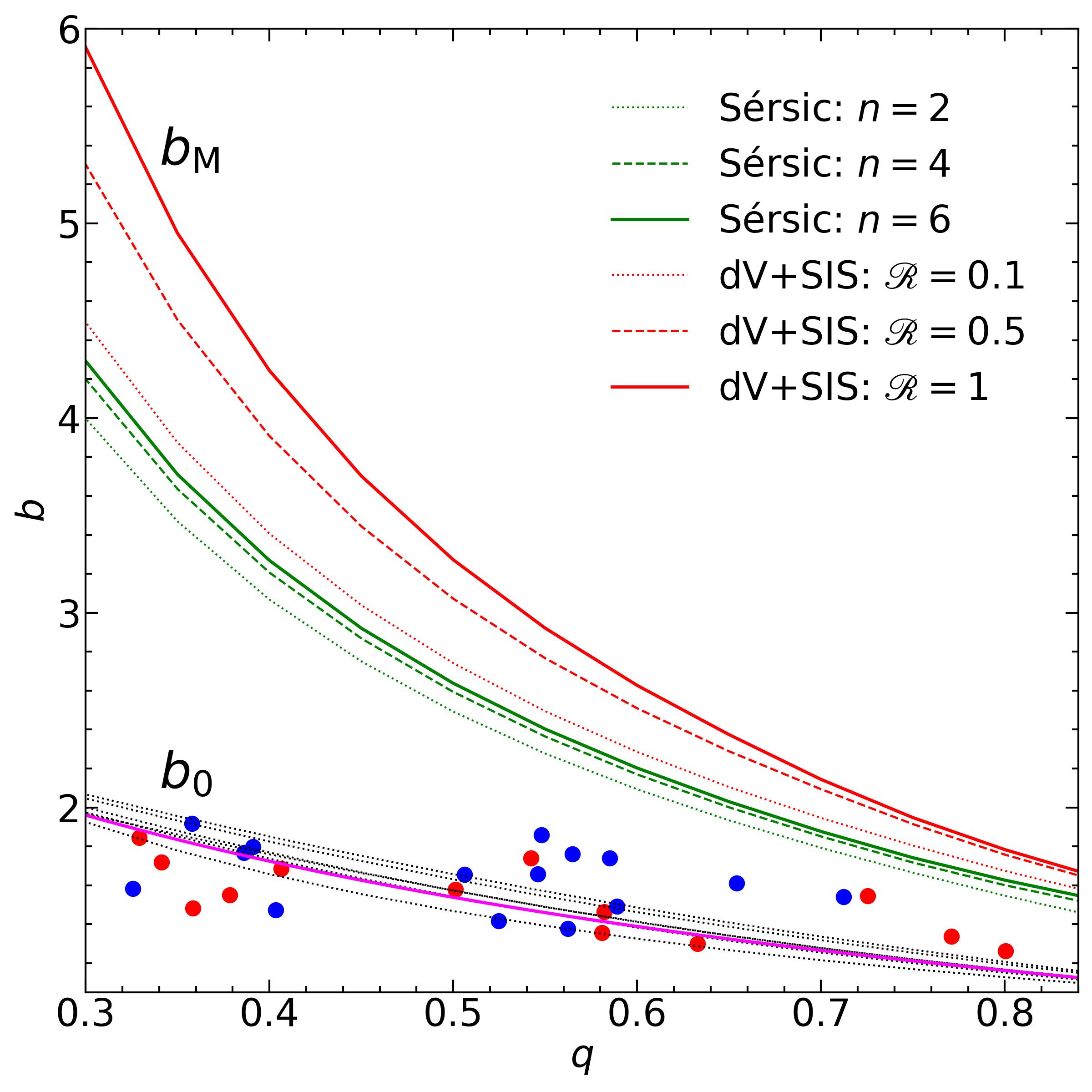}
        \caption{Mass-weighted average $\avb$ for the TNG50 galaxies in Table \ref{table:pmore}. The magenta line represents the empirically derived upper-limit $b\approx 1/(0.3+0.7\,q)$ for real galaxies, while the black dotted lines represent  the critical $\bz(q)$ relation for Sersic models of different $n$, and for de Vaucouleurs (dV) models in a SIS DM halo of different ${\cal R}\equiv M_{\rm SIS}(R_{\rm e})/\Mstar$. The green and red lines show the critical curves $\bM(q)$ for the same models.}
        \label{fig:figure14}
    \end{figure}

We finally turn to examine whether the $b$ parameter of the TNG50 ETGs behaves consistently with what found in recent modeling and comparison with observations. In Paper II it was found that for oblate ellipsoidal models, more flattened galaxies can support a larger amount of $b$-anisotropy than more round galaxies while maintaining $\vphiqm > 0$. In practice, the consistency condition for $\vphiqm > 0$ is $b\leq\bM(q)$, with $\bM(q)$ weakly dependent on the galaxy density profile and on the presence of a DM halo. Moreover, for galaxies in the SAURON integral-field survey an empirical limit on the anisotropy parameter $b$ as a function of flattening was found (Cappellari et al. 2007), with the majority of regular rotators well modeled by $b\lesssim 1/(0.3+0.7q)$; such limit has been confirmed by JAM in more recent surveys (Cappellari 2016). An interpretation of this limit was also proposed in Paper II: in fact it was there recognized that if $\gamma\approx 0$ (i.e., $\sigR\approx\sigphi$) from equation (\ref{eq:gamma_def}) necessarily $\Deltac >0$, and this requires $b\leq\bz(q)$ in case of constant $b$-anisotropy. Surprisingly, the $\bz(q)$ lines of the models were found almost superimposed to the empirical limit, supporting the idea that real galaxies are well described by $\sigR\approx\sigphi$, and so $\Deltac >0$ (see also Wang et al. 2021). It is therefore interesting to check how the selected TNG50 galaxies behave with respect to the two limits $\bM(q)$ and $\bz(q)$. In Figure \ref{fig:figure14} the values of $\avb$ and $q$ of the galaxies are plotted (blu points, Satellites; red points Central), together with the empirical $b(q)$ relation (the magenta line). For reference, some critical lines derived from analytical models are also presented. The black dotted lines near the magenta line are the $\bz(q)$ values for the ellipsoidal Sersic (1968) models with different profile index $n$, and for the ellipsoidal de Vaucouleurs (1948) models in a Singular Isothermal Sphere (SIS) DM halo, with different $\mathscr{R}=M_{\rm SIS}(R_{\rm e})/\Mstar$, where $M_{\rm SIS}(R_{\rm e})$ is the DM mass inside the effective radius of the stellar distribution (see equation [27] in Paper II). The upper red and green curves are the $\bM(q)$ values for the same models. The TNG50 galaxies reassuringly present $\avb$ values well below the consistency limit $\bM(q)$, and close to the magenta and $\bz(q)$ lines. From Figure \ref{fig:figure1} it is apparent that $\avg\approx 0$ for a significant fraction of the systems, so that from equation (7) with $\gamma=0$, 
$\avb\approx\bz(q)$ corresponds to the condition $\vphim^2 =0$: considering that $\avb$ is a mass-weighted value, and that $b$ is not constant for the TNG50 galaxies, it is quite surprising at the end how well the numerical galaxies respect the upper limit given by the empirically determined magenta line.

\section{Discussion and conclusions}

In this paper we studied in detail the internal dynamics of a selected sample of axysimmetric galaxies extracted from the highy-resolution cosmological simulation TNG50, and identified as ETGs at redshift zero. The aim was to check how well the selected galaxies are described by the axisymmetric JEs in equation (\ref{eq:3int_jeans}), and how the properties of their kinematical fields resemble observational results and theoretical expectations. From an initial sample of 81 bona-fide ETGs ($10^{10.5}\le\Mstar/\Msun\le 10^{12}$, specific star formation rates ${\rm sSFR} \le 10^{-12}$) in the TNG50 simulation, we restricted to a final sample of 29 highly axisymmetric systems. For these systems we recovered all the relevant kinematical fields (streaming velocities, velocity dispersion tensor) and the gravitational (stars plus DM) field. We then studied in detail the properties of the $b$-anisotropy parameter and other kinematical decomposition parameters of azimuthal motions, such as the Satoh' $k$, 
$\gamma$ in equation (\ref{eq:gamma_def}), and $\lambda$ in equation (\ref{eq:lambda_def}); finally we searched for systematic trends of these parameters as a function of galaxy internal and/or environmental properties. The main results can be summarized as follows.

\begin{itemize}

\item In the numerically well-sampled regions, the only significantly non-zero component of the streaming velocity is the rotational velocity $\vphim$, while $\vRm\approx\vzm\approx 0$ for all the 29 galaxies. Moreover, the off-diagonal components of the velocity dispersion tensor are very small when compared with the diagonal components, i.e., $\vRvzm\approx\vRvphim\approx\vzvphim\approx 0$, and the velocity dispersion tensor is cylindrically aligned, i.e., the hypotheses for the applicability of the JEs in equation (\ref{eq:3int_jeans}) are verified.

\item For each galaxy the l.h.s. and r.h.s. of the vertical and radial JEs were numerically constructed from the simulation data, and then compared. Over the well sampled regions of the galaxies the percentual differences of the two sides of the equations are small (of the order of 20\% or less), especially when considering the several steps in the procedure, involving numerical derivatives of numerically averaged quantities. This test, in addition to the indirect indication given by the strong structural axisymmetry, further confirms that the 29 selected galaxies are systems at equilibrium. 

\item For the 29 galaxies (15 satellite galaxies, and 14 central galaxies) the velocity dispersions $\sigR(R,z)$, $\sigz(R,z)$, and $\sigphi(R,z)$, and the functions 
$\Deltac (R,z)=\vphiqm -\sigR^2$ and $\Delta(R,z)=\vphiqm -\sigz^2$ were obtained from the simulation data. No significant qualitative differences are evident between the two groups of galaxies and, as expected, the values of the kinematic fields increase with increasing galaxy mass. 

\item $\Deltac (R,z)$ and $\Delta(R,z)$ in general present negative hourglass-shaped regions in the vicinity of the symmetry axis, while they are found positive on the equatorial regions. Therefore the Satoh decomposition cannot be used to describe the internal dynamics of TNG50 ETGs. The alternative kinematical decompositions described by the functions
$\gamma(R,z)=1-\sigphi^2/\sigR^2$ and $\lambda^2(R,z)=\vphim^2/\vphiqm$ were also considered, and the two functions were found non constant over the galaxies, but with common features, revealing a regular kinematical structure. Interestingly, the distribution of values of the mass-weighted $\avg$ peaks around $0$, as found for regular rotators with the JAM modeling (Cappellari 2016).

\item The $b(R,z) = \sigR^2/\sigz^2$ field is also not constant, and the studied galaxies are therefore not two-integral systems. In general the larger values of $b$ are along the equatorial plane, with maximum values near the center, but even along the equatorial plane the profile is not uniform, with values that tend to decrease from the center outward, and in many cases increase again farther out. In general, TNG50 galaxies with a more flattened stellar density distribution are characterized by higher values of $b$, whose values follow quite well the empirical upper limit $b(q)$ derived for real galaxies  (Cappellari et al. 2007, 2016). Overall, no significant dependence of $b$ on mass has been found, while the density flattening and at some level the environment (Satellite or Central) seem to play a more detectable effect. A phenomenological formula sufficiently flexible to capture the main properties of the $b(R,z)$ fields, that could be used in the modelization of observational data, has been proposed.

\end{itemize}

In conclusion, the selected axysimmetric and equilibrium TNG50 ETGs present kinematical fields characterized by clearly defined general properties. Perhaps one of the most interesting results, also for its implications in the Jeans based modelizations, is that the decomposition of azimuthal motions in ordered and velocity dispersion is not described by Satoh decomposition. The galaxies present a clear $b$-anisotropy, but the parameter $b$ is not constant over the galaxies, nor plane stratified with $b=b(z)$ as studied recently in Paper I and II; however the mass-averaged $\avb$ cluster near the upper limit $b(q)$ derived for real galaxies. It would be of great interest to study galaxies obtained from cosmological simulations of even higher resolution, to see if the trends here observed are confirmed, how the dynamics changes in the outer regions of the models (where the present numerical resolution is not high enough to guarantee robust results), and finally how the formation history of the simulated galaxies affects the features of the $b$, $\gamma$, and $\lambda$ fields.

\begin{acknowledgements}
An anonymous Referee is thanked for useful comments that improved the paper content and presentation. The IllustrisTNG simulations were undertaken with compute time awarded by the Gauss Centre for Supercomputing (GCS) under GCS Large-Scale Projects GCS-ILLU and GCS-DWAR on the GCS share of the supercomputer Hazel Hen at the High Performance Computing Center Stuttgart (HLRS), as well as on the machines of the Max Planck Computing and Data Facility (MPCDF) in Garching, Germany. The IllustrisTNG data used in this work are publicly available at https://www.tng-project.org/. L.D.D. acknowledges the hospitality of the Sterrenkundig Observatorium of Gent University during the visiting period of a Marco Polo Programme of the University of Bologna.
\end{acknowledgements}

\onecolumn
\begin{appendix}

\section{Table and Figures}

\begin{table}[ht!]
\caption{Properties of the 29 TNG50 galaxies in the final sample.}    
\label{table:pmore}     
\centering                          
\begin{tabular}{c  c  c  c  c  c  c  c  c  c  c c} 
\hline\hline              
\rule{0pt}{2.5ex} ID & $\log(\Mstar/\Msun)$ & $\rh$ & $b/a$ & $q$ & $\avb$ & $\sR$ & $\sz$ & $\avg$ & $\sqrt{\avl}$ & Richness & C/S \\[2.5pt] 
\hline\\
313694 & 10.6 & 3.54 & 0.984 & 0.386 & 1.79 & $-0.290$ & 0.246 & 0.271 & 0.788 & 33 & S\\
117259 & 10.7 & 5.96 & 0.978 & 0.358 & 1.96 &      0.040 & 0.013 & 0.117 & 0.750 & 160 & S\\
184937 & 10.8 & 2.20 & 0.999 & 0.712 & 1.62 &      0.264 & 0.126 & 0.084 & 0.566 & 103 & S\\
000010 & 10.8 & 6.41 & 0.998 & 0.327 & 1.77 & $-0.136$ & 0.293 & 0.212 & 0.854 & 390 & S\\
000009 & 10.8 & 5.47 & 0.993 & 0.506 & 1.85 & $-0.694$ & 0.329 & 0.162 & 0.706 & 390 & S\\
000005 & 10.9 & 7.24 & 0.988 & 0.565 & 1.99 & $-0.977$ & 0.253 & 0.261 & 0.633 & 390 & S\\
063874 & 10.9 & 3.40 & 0.988 & 0.589 & 1.53 & 0.078 & 0.222 & $-0.074$ & 0.628 & 209 & S\\
096763 & 10.9 & 4.72 & 0.954 & 0.548 & 2.01 & $-0.085$ & 0.320 & 0.094 & 0.593 & 141 & S\\
096764 & 11.0 & 5.26 & 0.956 & 0.585 & 1.84 & $-0.197$ & 0.257 & 0.114 & 0.577 & 141 & S\\
264885 & 11.1 & 3.53 & 0.978 & 0.546 & 1.70 &      0.084 & 0.111 & 0.169 & 0.597 & 47 & S\\
184932 & 11.2 & 9.09 & 0.979 & 0.391 & 1.88 & $-0.209$ & 0.113 & 0.113 & 0.708 & 103 & S\\
000002 & 11.3 & 7.69 & 0.954 & 0.654 & 1.84 & $-0.622$ & 0.334 & 0.066 & 0.474 & 390 & S\\
117252 & 11.3 & 7.72 & 0.992 & 0.403 & 1.58 & $-0.405$ & 0.267 & 0.203 & 0.806 & 160 & S\\
143883 & 11.3 & 4.56 & 0.986 & 0.525 & 1.50 & $-0.043$ & 0.262 & $-1.02$ & 0.490 & 129 & S\\
000001 & 11.6 & 8.43 & 0.997 & 0.562 & 1.48 & $-0.297$ & 0.203 & 0.095    & 0.655 & 390 & S\\\\
\hline\\
563732 & 10.6 & 5.29 & 0.985 & 0.329 & 1.85 & $-0.015$ & 0.103 & 0.130 & 0.653 & 2 & C\\
572328 & 10.6 & 4.18 & 0.979 & 0.358 & 1.59 & $-0.306$ & 0.184 & 0.203 & 0.855 & 2 & C\\
546114 & 10.7 & 3.10 & 0.960 & 0.501 & 1.63 & 0.069 & 0.244 & $-0.249$ & 0.718 & 2 & C\\
516101 & 10.7 & 3.25 & 0.973 & 0.582 & 1.56 & $-0.241$ & 0.181 & $-0.115$ & 0.563 & 4 & C\\
478216 & 10.8 & 5.04 & 0.976 & 0.378 & 1.67 & $-0.419$ & 0.250 & $-0.026$ & 0.786 & 8 & C\\
543114 & 10.8 & 3.09 & 0.951 & 0.542 & 1.72 & 0.517 & 0.296 & $-1.83$ & 0.140  & 1 & C\\
454171 & 10.9 & 5.24 & 0.987 & 0.294 & 1.90 & $-0.372$ & 0.286 & 0.231 & 0.797 & 5 & C\\
428177 & 11.0 & 4.68 & 0.993 & 0.581 & 1.41 & 0.238 & 0.214 & $-0.271$ & 0.620 & 12 & C\\
498522 & 11.0 & 6.81 & 0.984 & 0.407 & 1.83 & $-0.491$ & 0.082 & 0.139 & 0.748 & 3 & C\\
450916 & 11.1 & 7.64 & 0.973 & 0.725 & 1.63 & $-0.301$ & 0.199 & $-0.112$ & 0.073 & 5 & C\\
482155 & 11.1 & 6.58 & 0.969 & 0.341 & 1.81 & $-0.385$ & 0.289 & 0.050 & 0.779 & 3 & C\\
418335 & 11.2 & 8.98 & 0.991 & 0.800 & 1.28 & $-0.022$ & 0.067 & 0.033 & 0.464 & 9 & C\\
390932 & 11.4 & 9.97 & 0.997 & 0.633 & 1.36 & $-0.333$ & 0.106 & $-0.219$ & 0.547 & 8 & C\\
319730 & 11.8 & 11.8 & 0.979 & 0.771 & 1.40 & $-0.345$ & 0.128 & 0.101 & 0.518 & 18 & C\\\\
\hline                              
\end{tabular}
\tablefoot{From left to right: ID of the galaxy, total stellar mass $\Mstar$, stellar (volumic) half-mass radius $\rh$ in kpc, intermediate-to-long and short-to-long ($q$) axial ratios, mass-weighted average $\avb$ and its associated shape parameters $\sR$ and $\sz$ (see Section 4), mass-weighted average $\avg$ and $\sqrt{\avl}$, number of galaxies inside the parent halo, relative position of the galaxy inside the parent halo (Central or Satellite).}
\end{table}

    \begin{figure}[ht!]
       \centering
       \includegraphics[width=0.9\linewidth]{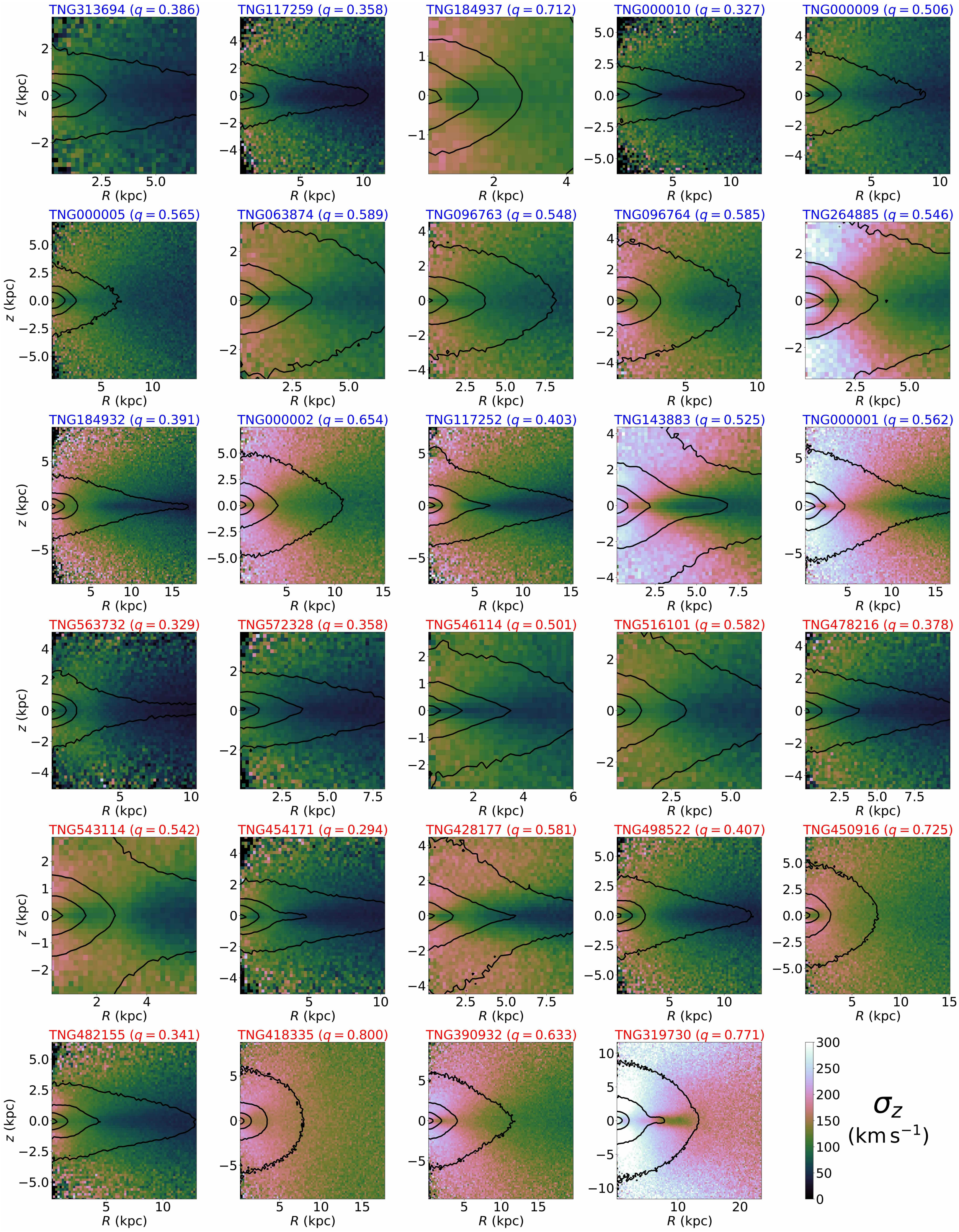}
       \caption{$\sigz (R,z)$ field of the $29$ galaxies in Table \ref{table:pmore}, over the region $R\leq 2\rh$, $\vert z\vert\leq\rh$. The galaxies are divided into satellite (S, blue labels) and central (C, red labels); in each of the two groups, $\Mstar$ increases from top-left to bottom-right. Black lines are the contours of the three-dimensional stellar density $\rhos(R,z)$, with the value of the innermost contour of $10^7\Msun {\rm kpc}^{-3}$, decreasing by a factor of 10 between two successive contours. For comparison the axial ratio $q$, computed from the second-order mass tensor ${\cal S}$ in equation (\ref{eq:shape_tens}), is also given above each map.}
        \label{fig:figure3}
    \end{figure}

    \begin{figure}[ht!]
       \centering
       \includegraphics[width=0.9\linewidth]{figure4}
       \caption{$\sigR (R,z)$ field of the $29$ galaxies in Table \ref{table:pmore}. For details see the caption of Figure \ref{fig:figure3}.}
        \label{fig:figure4}
    \end{figure}

    \begin{figure}[ht!]
       \centering
       \includegraphics[width=0.9\linewidth]{figure5.jpg}
       \caption{$\sigphi (R,z)$ field of the $29$ galaxies in Table \ref{table:pmore}. For details see the caption of Figure \ref{fig:figure3}.}
        \label{fig:figure5}
    \end{figure}

    \begin{figure}[ht!]
                \centering
            \includegraphics[width=0.95\linewidth]{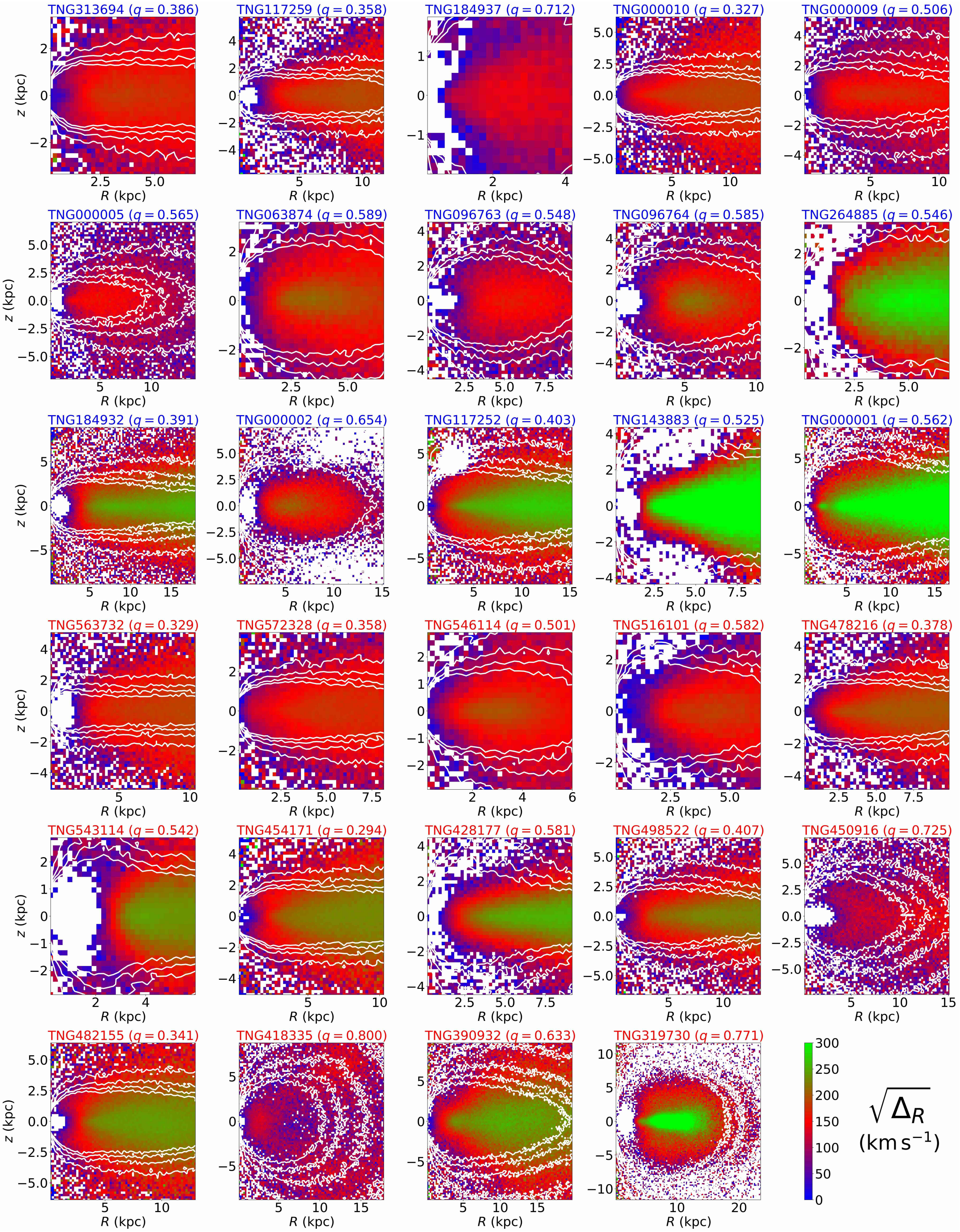}
            \caption{$\sqrt{\Deltac (R,z)}$ field for the 29 galaxies in Table \ref{table:pmore}; white regions correspond to negative values of $\Deltac (R,z)$. The white lines bound the regions with a number of stellar particles per pixel respectively larger than 200, 150, 100, and 50.}
             \label{fig:figure6}
        \end{figure}

    \begin{figure}[ht!]
                \centering
            \includegraphics[width=0.95\linewidth]{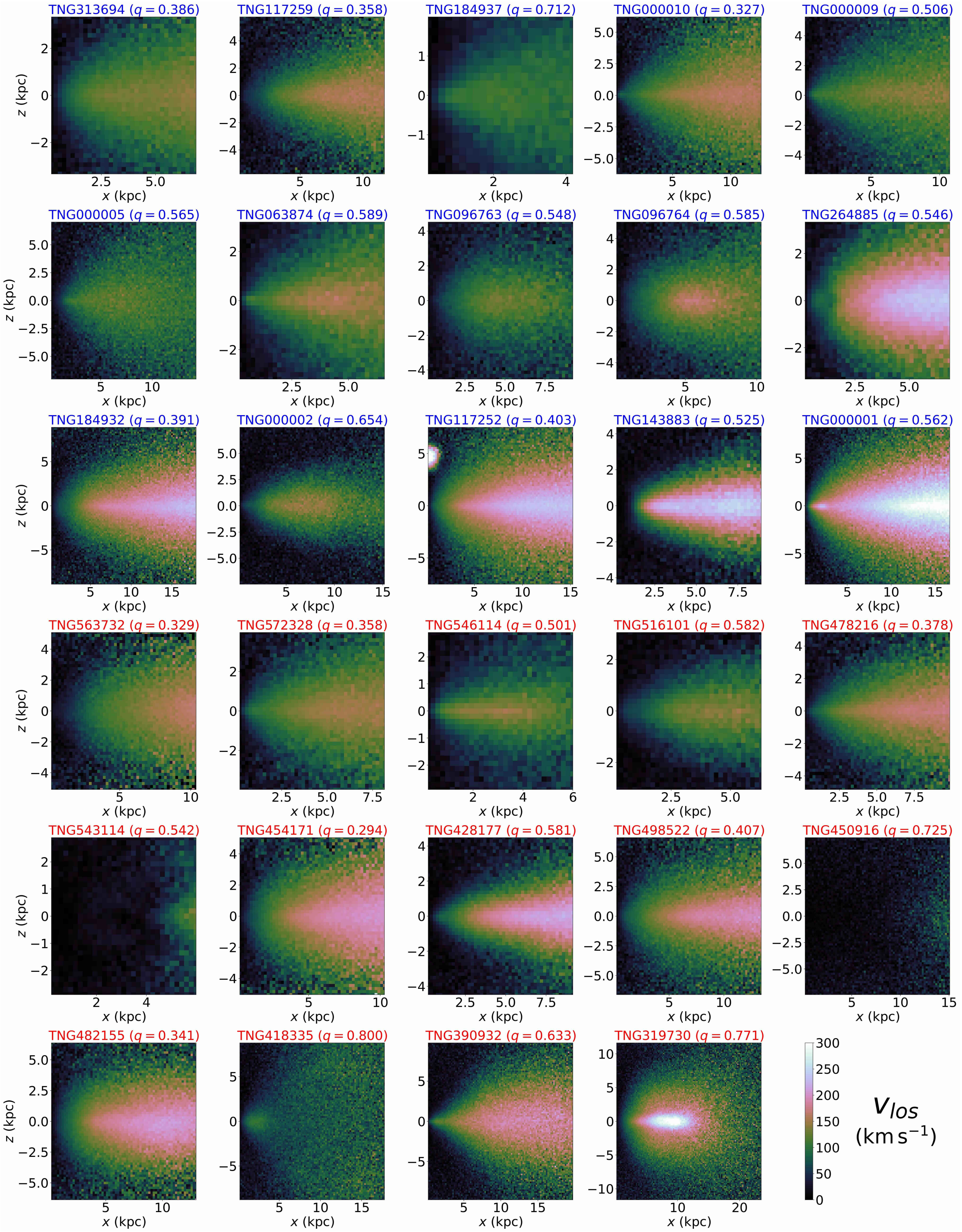}
            \caption{The line-of-sight projected streaming velocity field $v_{\rm los}(R,z)$ for the 29 galaxies in Table 1, computed along the $y$-axis; for the two galaxies TNG000010 and TNG390932 a comparison with the intrinsic streaming velocity field can be obtained from inspection of Figure \ref{fig:figure2}.}
             \label{fig:figure7}
        \end{figure}

    \begin{figure}[ht!]
                \centering
            \includegraphics[width=0.95\linewidth]{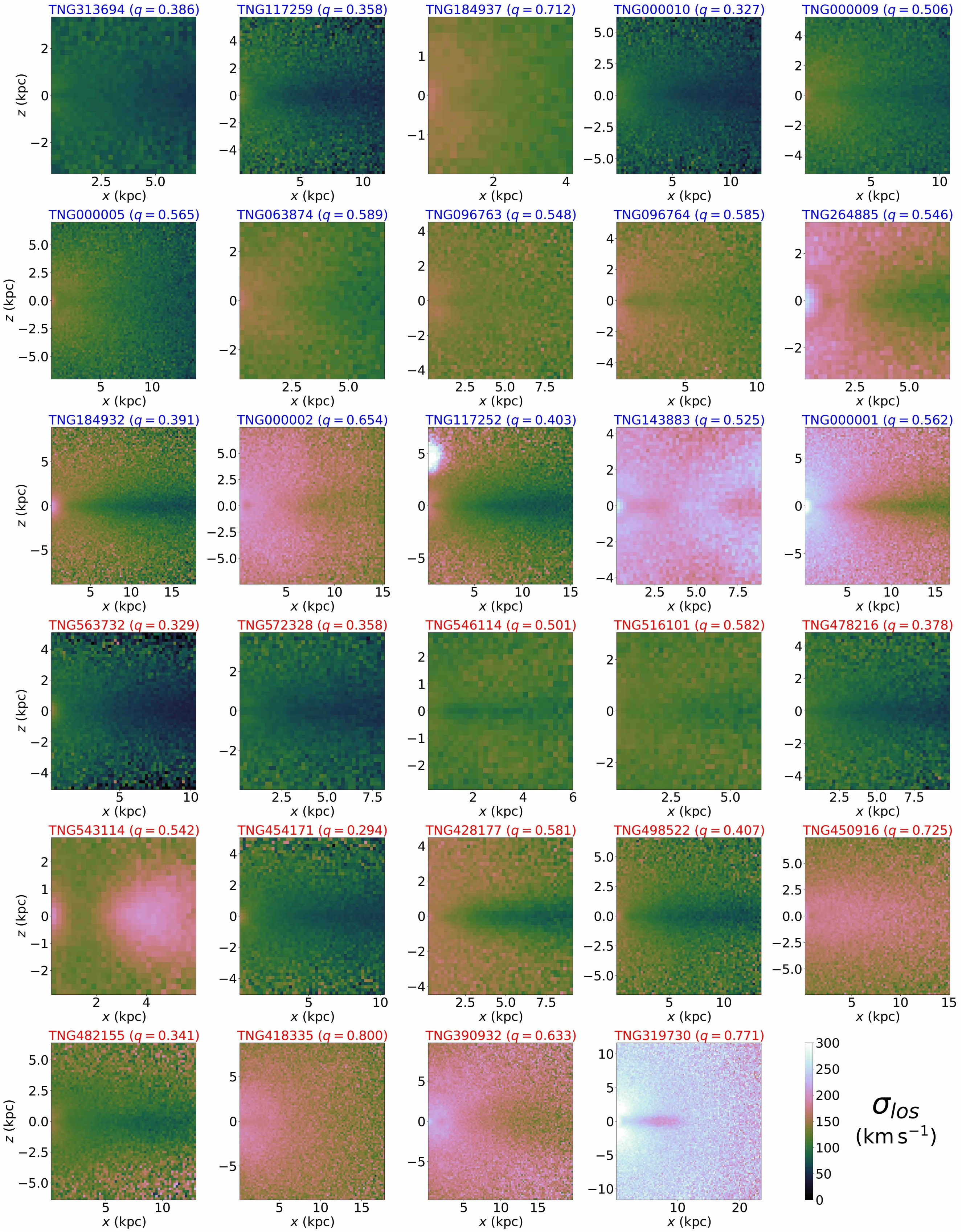}
            \caption{The line-of-sight projected velocity dispersion field $\sigma_{\rm los}(R,z)$ for the 29 galaxies in Table \ref{table:pmore}, computed along the $y$-axis.}
             \label{fig:figure8}
        \end{figure}
        
    \begin{figure}[ht!]
                \centering
            \includegraphics[width=0.95\linewidth]{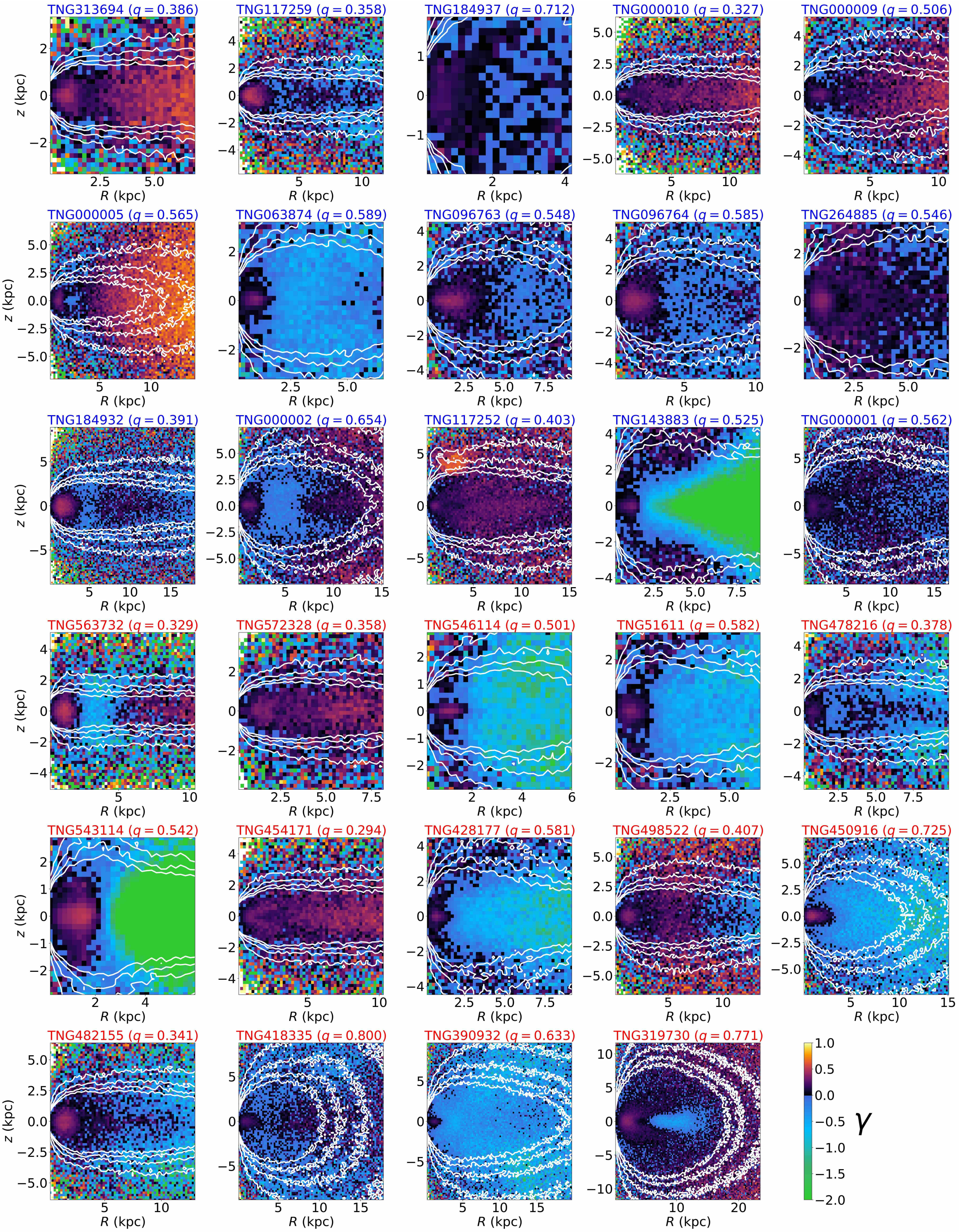}
            \caption{The decomposition parameter $\gamma(R,z)$ in equation (\ref{eq:gamma_def}).
            Blue-green regions correspond to negative values, i.e., $\sigphi\geq\sigR$, while in red-orange regions $\gamma$ is positive, with $\sigphi\leq\sigR$. For the meaning of the white lines, see the caption of Figure \ref{fig:figure6}.}
             \label{fig:figure9}
        \end{figure}

    \begin{figure}[ht!]
            \centering
            \includegraphics[width=0.95\linewidth]{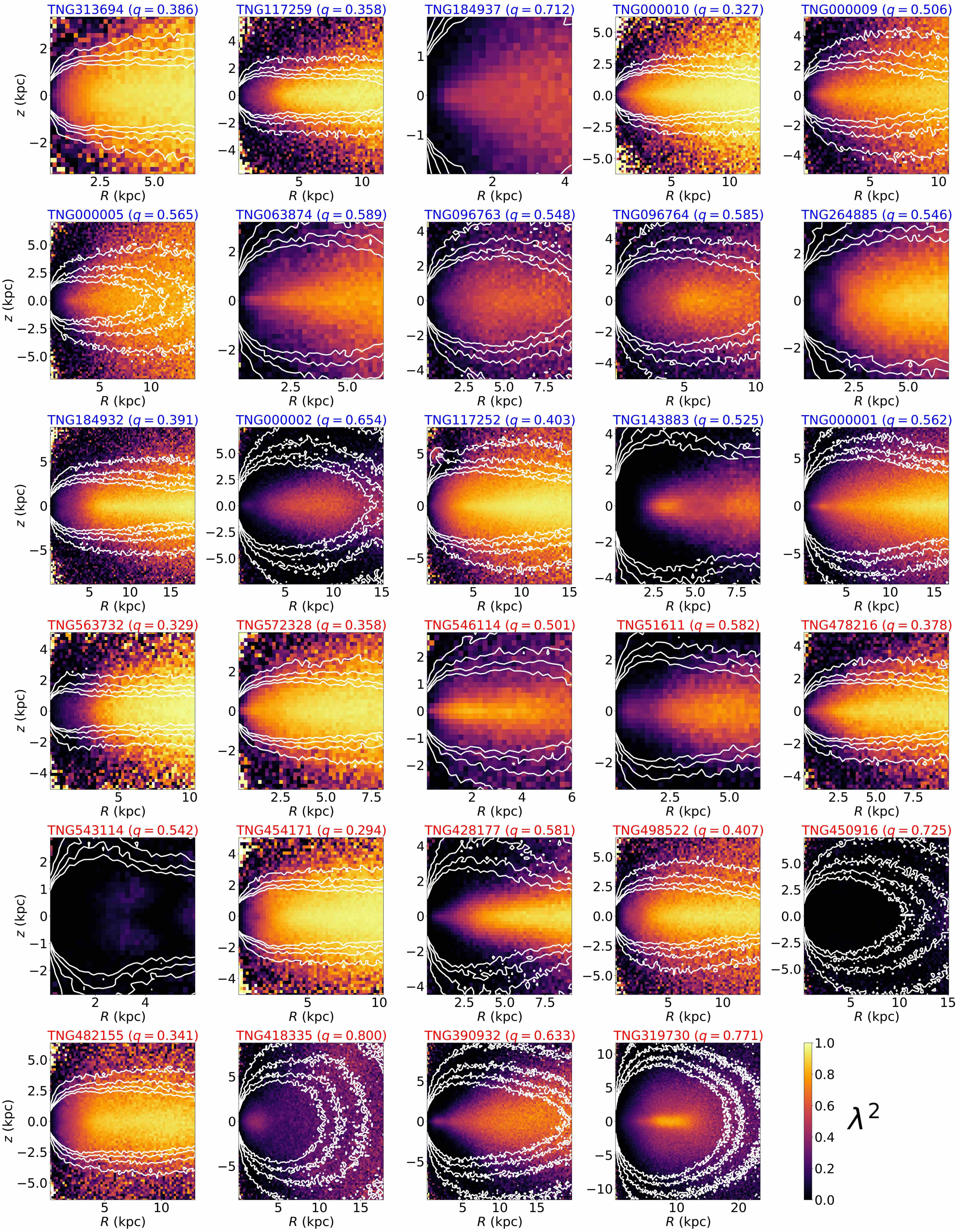}
            \caption{Maps of the decomposition parameter $\lambda^2(R,z)$ in equation (\ref{eq:lambda_def}). For the meaning of the white lines, see the caption of Figure \ref{fig:figure6}.}
            \label{fig:figure10}
        \end{figure}     
        
        \begin{figure}[ht!]
            \centering
            \includegraphics[width=0.95\linewidth]{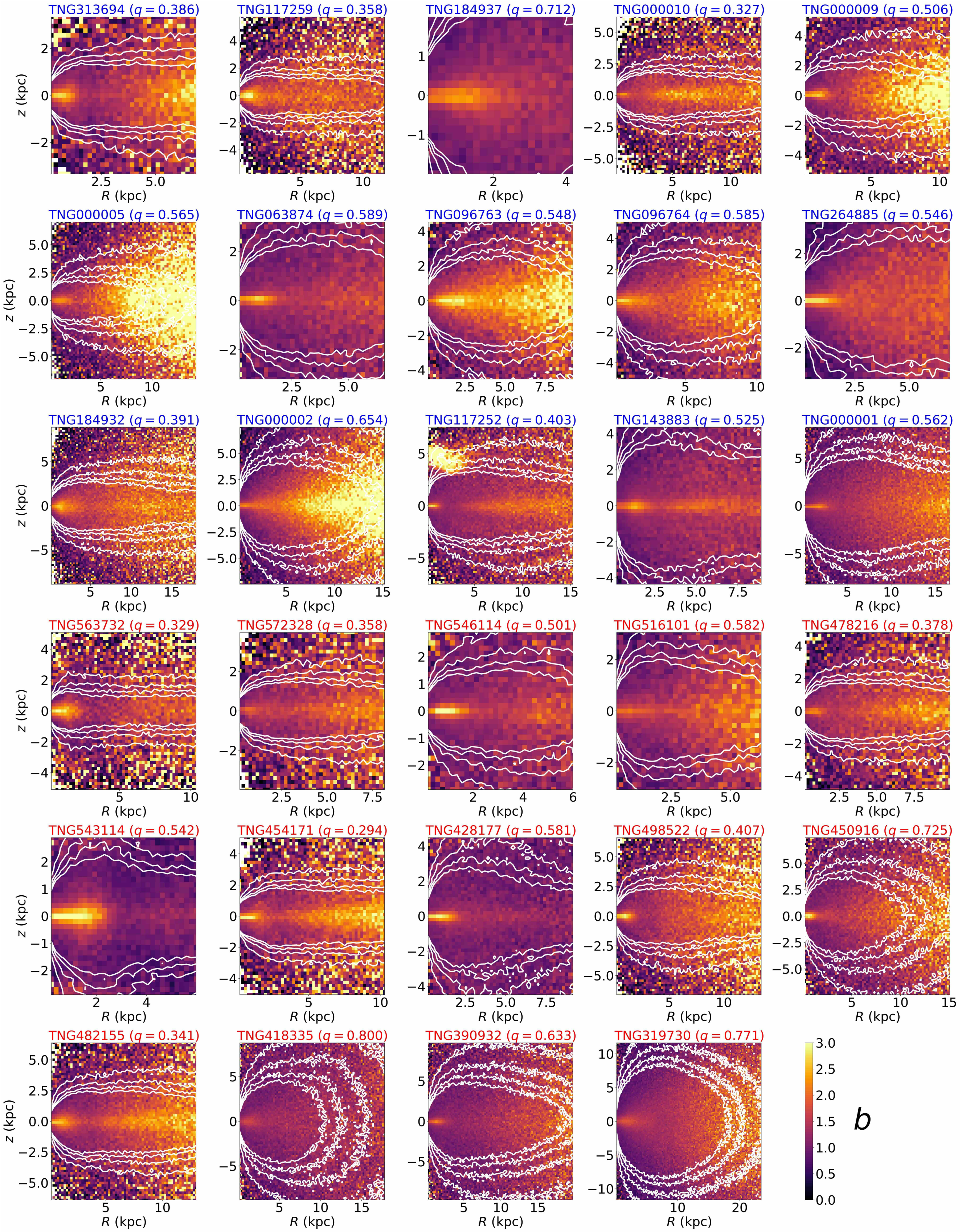}
            \caption{Maps of the anisotropy parameter $b = \sigR^2/\sigz^2$.
            For the meaning of the white lines, see the caption of Figure \ref{fig:figure6}.}
            \label{fig:figure11}
        \end{figure}   

\end{appendix}
\end{document}